\documentclass{ieeeaccess}
\usepackage{cite}
\usepackage{amsmath,amssymb,amsfonts}
\usepackage{algorithmic}
\usepackage{graphicx}
\usepackage{subcaption}
\usepackage{textcomp}
\usepackage{amsmath}
\usepackage{caption,setspace}
\usepackage{xcolor, soul}
\usepackage{hyperref} 
\usepackage{etoolbox}

\def\thevol{8}
\def\theyear{2020}
\makeatletter
\patchcmd{\@evenfoot}{2016}{\theyear}{}{}
\patchcmd{\@oddfoot}{2016}{\theyear}{}{}
\makeatother
    
\def\BibTeX{{\rm B\kern-.05em{\sc i\kern-.025em b}\kern-.08em
    T\kern-.1667em\lower.7ex\hbox{E}\kern-.125emX}}   
    
\begin{document}

\history{This is the author’s version of an article that has been published as: \textbf{O. Amador, I. Soto, M. Calderón and M. Urueña, "Experimental Evaluation of the ETSI DCC Adaptive Approach and Related Algorithms," in IEEE Access, vol. 8, pp. 49798-49811, 2020.}
This work is licensed under a Creative Commons Attribution 4.0 License. For more information, see https://creativecommons.org/licenses/by/4.0/.}

\doi{of final version of record: \href{http://dx.doi.org/10.1109/ACCESS.2020.2980377}{10.1109/ACCESS.2020.2980377}}

\title{Experimental Evaluation of the ETSI DCC Adaptive Approach and Related Algorithms}
\author{\uppercase{Oscar Amador}\authorrefmark{1},
\uppercase{Ignacio Soto\authorrefmark{1}, Maria Calderón\authorrefmark{1}, and Manuel Urueña\authorrefmark{2}}
}
\address[1]{Departamento de Ingeniería Telemática, Universidad Carlos III de Madrid, 28911 Leganés, Spain (e-mail: oamador@pa.uc3m.es; isoto@it.uc3m.es; maria@it.uc3m.es).}
\address[2]{Escuela Superior de Ingenieros y Tecnología, Universidad Internacional de la Rioja, 26006 Logroño, Spain (e-mail: manuel.uruena@unir.net).}
\tfootnote{This work was partially supported by the Spanish Ministerio de Economía y Competitividad through the Texeo project (TEC2016-80339-R).}

\markboth
{Amador \headeretal: Experimental Evaluation of the ETSI DCC Adaptive Approach and Related Algorithms}
{Amador \headeretal: Experimental Evaluation of the ETSI DCC Adaptive Approach and Related Algorithms}

\corresp{Corresponding author: Oscar Amador (e-mail: oamador@pa.uc3m.es).}

\begin{abstract}
Decentralized Congestion Control (DCC) mechanisms have been a core part of protocol stacks for vehicular networks since their inception and standardization. The ETSI ITS-G5 protocol stack for vehicular communications considers the usage of DCC not only in the network or access layers, but also as a part of the cross-layer architecture that influences how often messages are generated and transmitted. ETSI DCC mechanisms have evolved from a reactive approach based on a finite state machine, to an adaptive approach that relies on a linear control algorithm. This linear control algorithm, called LIMERIC, is the basis of the mechanism used in the ETSI DCC Adaptive Approach. The behavior of this algorithm depends on a set of parameters. Different values for these parameters have been proposed in the literature, including those defined in the ETSI specification. A recent proposal is Dual-$\alpha$, which chooses parameters to improve convergence and fairness \textcolor{black}{when the algorithm has to react to fast changes in the use of the shared medium (transitory situations)}. This article evaluates, by means of simulations, the performance of the ETSI DCC Adaptive Approach and related algorithms, considering both steady state and transitory situations. Results show that a bad selection of parameters  can make a DCC algorithm ineffective, that the ETSI DCC Adaptive algorithm performs well in steady state conditions, and that Dual-$\alpha$ performs as well in steady state conditions and outperforms the ETSI DCC Adaptive Approach in transitory scenarios.
\end{abstract}

\begin{keywords}
Congestion control, Convergence, Cooperative communication, ETSI, Vehicular Ad Hoc Networks.
\end{keywords}

\titlepgskip=-15pt

\maketitle

\section{Introduction}
\label{sec:introduction}

Network resources such as bandwidth are limited and subject to conditions that can vary from one moment to another. In the case of Vehicular Ad-Hoc Networks (VANETs), conditions can vary in many ways, e.g., one car can change from driving in a one-way street to merge into a busy boulevard, and this implies changes in the speed of the car and the way it interacts with other road users. In the world of Cooperative, Connected and Automated Mobility (CCAM), these changes are also reflected in the behavior of the vehicles’ On-Board Units. For example, in the Cooperative Awareness \textcolor{black}{(CA)} service, the generation frequency of its message (CAM --- Cooperative Awareness Message) as specified by the European Telecommunications Standards Institute (ETSI) depends on vehicle dynamics (i.e., speed, acceleration, and heading) --- the more they change, more messages are created \cite{etsiCAM}. Also, there are possible scenarios where a high number of \textcolor{black}{vehicles} (or more generally, Intelligent Transport Systems Stations) are within the range of each other (e.g., a traffic jam, a parking lot, etc.), which could also cause network congestion (i.e., a degradation in the network performance due to it transmitting more data than it can handle). With a typical CAM message size nearing 400 bytes at a frequency ranging from 1 to 10 Hz, events where messages can overload the 6 Mbit/s channel in the 5.9 GHz are likely, even without considering that multiple services have to coexist also in the medium.

If a vehicular network reaches a point where it is congested, messages are less likely to be received.  For the case of the Cooperative Awareness service, this means that a given vehicle has less information regarding its surroundings (e.g., knowing the number and position of other cars), and if other services are considered --- such as the Decentralized Environmental Notification service, which alerts of hazardous situations --- this could imply a higher risk for the road users. Also, a network that is not congested is able to better allocate network resources fairly to \textcolor{black}{vehicle}s, which could reflect on the quality and the quantity of information \cite{dataVANETs} vehicles base their decisions on.

Strategies for avoiding network congestion have been adopted in the field of vehicular networks. The ETSI ITS-G5 protocol stack considers the use of Decentralized Congestion Control (DCC) mechanisms that apply on multiple layers\cite{etsiNewDcc}, as do the standards established by the Society of Automotive Engineers (SAE International)\cite{saeDcc}. These mechanisms try different approaches to DCC that range from limiting the rate at which \textcolor{black}{vehicle}s generate and send messages, to limiting the power which \textcolor{black}{vehicle}s use to transmit these messages, or a combination of both.

\textcolor{black}{Our previous work \cite{Letter} analytically studies the performance of the ETSI DCC Adaptive Approach and proposes the \mbox{Dual-$\alpha$} algorithm to achieve better performance in transitory scenarios. We call transitory scenarios those in which changes in conditions significantly modify the share of the medium available to each \textcolor{black}{vehicle}; and we call steady state scenarios those in which the share of the medium that corresponds to each \textcolor{black}{vehicle} tends to a stable value that changes slowly with time. In \cite{Letter}, simulations are used to compare the performance of the ETSI DCC Adaptive Approach and \mbox{Dual-$\alpha$} in transitory scenarios. However, in these simulations, the \textcolor{black}{vehicle}s are static; only CAM traffic is present; and CAM message generation is at fixed intervals, independently of vehicle dynamics and feedback from DCC.} 

\textcolor{black}{The simulation-based work we present here} is the first one to perform a thorough evaluation of the ETSI DCC Adaptive Approach and compare it with related mechanisms considering: 
\begin{itemize}
	\item Steady state and transitory scenarios. \textcolor{black}{A DCC algorithm in steady state scenarios must maximize the use of the shared medium, while avoiding congestion; and, in transitory scenarios, the algorithm must be able to adapt as fast as possible to new situations.}   
	\item Realistic CAM generation based on vehicle dynamics and feedback from DCC.
	\item The effect of background traffic coexisting with CAM messages.
	\item A benchmark without using DCC.
\end{itemize}

This work is divided as follows. In Section \ref{sec:dcc}, congestion control mechanisms are explored, including not only the ones standardized by ETSI (reactive and adaptive approaches), but also the mechanism that introduced the adaptive approach --- LIMERIC \cite{LimericOriginal} --- and its updates proposed in \cite{LimericStability},  \cite{LimericForwarding}; as well as Dual-$\alpha$ \cite{Letter}, which tries to address some of the issues that arise because of the value of standardized parameters. Section \ref{sec:simdesc} describes the simulation setups (both steady state and transitory scenarios) where we evaluate these approaches and the metrics to be used. Section \ref{sec:simres} analyzes the results of these simulations. Finally, we conclude in section \ref{sec:conclusion}.

\section{Decentralized Congestion Control}
\label{sec:dcc}
Different Decentralized Congestion Control (DCC) mechanisms have been proposed and standardized for vehicular networks, in order to maintain network stability and fair resource allocation to comply with network requirements (e.g., reliability, fair channel access for all nodes, etc.). Most of the DCC mechanisms stem from the basic idea behind the concept of congestion control in a network: modulating the traffic that enters the network. This modulation can be achieved through different ways, such as controlling the number of messages a \textcolor{black}{vehicle} can send, the transmission power used to send messages, and the channel data rate. In the ETSI standards, solutions that control the number of messages a \textcolor{black}{vehicle} can send (TRC, for Transmit Rate Control) are the most relied upon \cite{etsiNewDcc}. In this type of mechanisms, each \mbox{\textcolor{black}{vehicle}} controls the rate at which it sends messages in order to get to an optimal rate allocation that is given by a metric called Channel Busy Ratio (CBR), which is the time-dependent value between zero and one representing the fraction of time that a radio channel is busy with transmissions. ETSI standards have defined these mechanisms and grouped them into reactive and adaptive approaches. The ETSI DCC architecture consists of components that are located on different layers within the ETSI vehicular network protocol stack: management, facilities, networking and transport, and access layers.

\begin{figure}[h]
	\centering
	\includegraphics[width=205px,height=200px,clip,keepaspectratio]{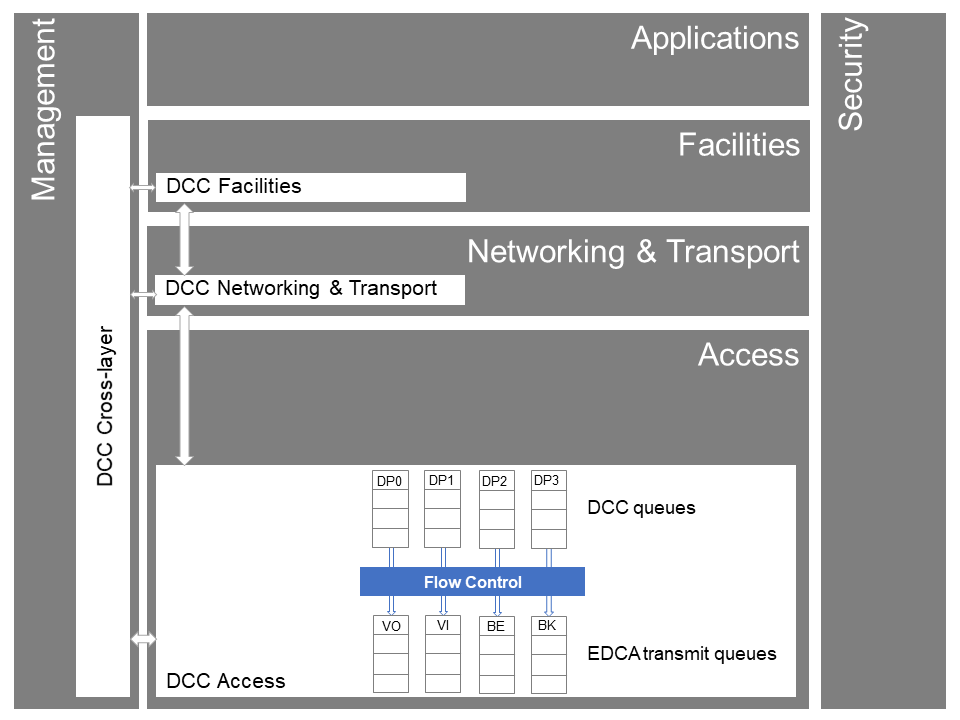}
	\caption{ETSI DCC Architecture}
	\label{fig:dccarchitecture}
\end{figure} 

Fig. \ref{fig:dccarchitecture} shows the interactions between components of the DCC mechanism in different layers. The DCC Access component senses the channel busy ratio (CBR) to which the congestion mechanisms react or adapt. A number of services in the facilities layer, such as the Cooperative Awareness basic service \cite{etsiCAM}, adapt their message generation frequency according to the rate DCC assigns them, while others such as the Decentralized Environmental Notification basic service do not consider DCC when generating their messages (DENM messages) due to the nature of the service itself. Messages generated by services at the facilities layer are assigned to data profiles (DP) with four different priorities: DP0 to DP3, with DP0 having the top priority (e.g., DENM and CAM messages correspond to data profiles DP0 and DP2 respectively). \textcolor{black}{These messages go down the protocol stack until they arrive at the access layer, where the DCC flow control mechanism limits the rate at which they are sent to the Enhanced Distribution Coordination Access (EDCA) transmit queues. The EDCA transmit queues are part of the medium access control mechanism \cite{etsiEDCA} used by the ETSI protocol stack. Packets waiting at the access layer due to the rate limitation of the DCC mechanism are stored in queues according to their profile (DP queues). When the DCC mechanism allows sending a packet to the EDCA queues, the packet sent is the first one in the highest priority DP queue with packets.} Data profiles DP0, DP1, DP2 and DP3 map to the EDCA Voice, Video, Best Effort and Background Access Categories (AC) respectively. These ACs have different properties to provide quality of service, with shorter arbitration inter-frame space (AIFS) periods for messages with higher priorities. This DCC architecture (Fig. \ref{fig:dccarchitecture}) is used by the two ETSI standardized congestion control approaches: reactive and adaptive.

\subsection{Reactive Approach}

The Reactive Approach is a DCC mechanism that was defined initially in ETSI TS 102 687 V1.1.1 in 2011 and updated in 2018 \cite{etsiNewDcc}. It is a DCC mechanism based on a Finite State Machine (FSM) with three main states: relaxed, active, and restrictive, each one having a CBR threshold to reach a neighboring state, since states can only be reached by a neighbor. Each state has values that limit parameters such as transmission power, rate, and data rate. The ETSI DCC Reactive Approach allows the control of these parameters either individually or combined with each other. While any of these techniques may be used, the current standard itself only provides an example of Transmission Rate Control (TRC).

This mechanism relies on measurements of the channel load and statistics on transmitted packets. If the CBR is not within the limits set for the current state, the state changes to a neighbor that is closer to the limits. The active state, according to the current standard, can be divided in $n$ sub-states, each one with its own threshold. The \textit{Relaxed} state, being the least restrictive, allows a transmission rate of a maximum of 20 messages per second as long as the CBR is below 0.30. The message allowance lowers each time an upper state is reached, with the \textit{Active} state ranging from 10 to 2 messages per second depending on how many sub-states exist and which one the system is on. The \textit{Restrictive} state allows transmitting at a rate of one message per second, which complies with the minimum requirements of the Cooperative Awareness Basic Service, but could affect traffic from other services, since the current standard does not specify different schedulers for the existing access categories.

While the reactive approach has been a long-standing standardized mechanism, several works have measured its performance and limitations \cite{fairness}. In \cite{dccFSMplat}, the authors demonstrate that the use of a FSM causes underutilization of the medium, and in \cite{dccHarri} the authors concur in the underutilization aspect. The reactive approach has also been compared to alternatives such as LIMERIC in \cite{LimericOriginal} and \cite{LimericStability}, where adaptive approaches outperformed the standardized reactive approach. 

\subsection{Adaptive Approach}

Adaptive approach mechanisms are basically linear control systems where the process variable is the transmission rate and the setpoint is a fraction of the channel capacity. Linear Message Rate Integrated Control (LIMERIC) \cite{LimericOriginal} is the base for the ETSI adaptive approach defined in the current standard. An adaptive congestion control approach has three objectives: 1) converging to a predefined channel utilization level; 2) achieving local fairness among immediate neighboring vehicles; and 3) achieving global fairness among all vehicles contributing to congestion. There have been several iterations of the LIMERIC algorithm proposed by the same authors, using several combinations of parameter values. In this section, parameters for LIMERIC from \cite{LimericOriginal}, \cite{LimericStability} and \cite{LimericForwarding} are described, along with those from the mechanism in the ETSI standard\cite{etsiNewDcc}. Finally, the DCC algorithm from \cite{Letter} is described, which is a proposed improvement to the standard \cite{etsiNewDcc}.

\paragraph{LIMERIC}
LIMERIC is a distributed and adaptive linear rate-control algorithm where each vehicle adapts its message rate in such a way that the total channel load converges to a specified target. First proposed in \cite{LimericOriginal}, it relies on the assumption that each \textcolor{black}{vehicle} can accurately measure the CBR at a given time. Each \textcolor{black}{vehicle} updates its fraction of medium usage allowance ($\delta$) for the instant $n$ using equation \ref{equation:limOrig}.  

\begin{equation}
\label{equation:limOrig}
\delta(n) = (1 - \alpha)\cdot\delta(n - 1) + \beta\cdot(CBR_{target} - CBR(n - 1))
\end{equation}

\textcolor{black}{where $\alpha$ and $\beta$ are parameters of the algorithm that influence convergence speed, stability, and achieved medium utilization in steady state (refer to  \cite{LimericOriginal}  for a detailed explanation). The second term of the right expression represents the fraction of channel occupancy the \textcolor{black}{vehicle} should yield or acquire ($\beta$ is a scaling factor). In the first term of the right expression, $\alpha$ assigns a weight to the previous $\delta$, to improve the ability of the system to react to changes. The steady state total utilization of the medium achieved by LIMERIC is given by equation \ref{equation:cbrlimeric}  \cite{LimericOriginal}:}

\begin{equation}
\label{equation:cbrlimeric}
CBR = \frac{K\cdot\beta\cdot CBR_{target}}{\alpha + K\cdot\beta}
\end{equation}

\textcolor{black}{where K is the number of vehicles sharing the medium.} Values for $\alpha$ and $\beta$ provided in the original proposal for LIMERIC \cite{LimericOriginal} are shown in Table \ref{table:limOrigPar}. Using equation \ref{equation:limOrig}, the proposed $\beta$ guarantees convergence for up to 285 vehicles \cite{LimericOriginal}. 

\begin{table}[h!]
	\centering
	\caption{Parameter values for LIMERIC from \cite{LimericOriginal}}
	\begin{tabular}{|c | c|}
		\hline
		\textbf{Parameter} & \textbf{Value} \\[0.5ex]
		\hline
		$\alpha$ & 0.1 \\
		$\beta$ & 0.0067 \\
		CBR Target & 0.60 \\[1ex]
		\hline
	\end{tabular}
	\label{table:limOrigPar}
\end{table}

To deal with scenarios that are very dense, where the values for $\alpha$ and $\beta$ cannot guarantee convergence, authors in \cite{LimericOriginal} propose a modification for the update formula (equation \ref{equation:limOrig}) that includes a threshold $X$ that limits the maximum variation, as shown in equation \ref{equation:limSigno}:

\begin{equation}
\label{equation:limSigno}
\begin{split}
&\delta(n) = (1 - \alpha)\cdot\delta(n - 1)\\&+ sign(CBR_{target} - CBR(n - 1))\\ &\cdot min[X,\beta\cdot abs(CBR_{target} - CBR(n - 1))]\end{split}
\end{equation}

This means that, if the magnitude of the updated offset given by the absolute value of $CBR_{target} - CBR(n - 1)$ multiplied by $\beta$ exceeds the value of $X$, the maximum offset would be $X$. The resulting offset is then added or subtracted from $(1 - \alpha)\cdot\delta(n - 1)$ depending on whether $CBR_{target} - CBR(n - 1)$ is positive or negative. This formula is the basis for the logic behind the ETSI Adaptive Approach, however, it was not mentioned in \cite{LimericStability} where LIMERIC authors propose a new set of parameters to deal with stability challenges in scenarios with high congestion, nor in \cite{LimericForwarding}, where they use parameter values closer to those in the ETSI standard.

\paragraph{Updates to LIMERIC}
The values proposed for $\alpha$ and $\beta$ have an effect on the time it takes for the system to converge \cite{LimericOriginal}, and the number of nodes it can handle in a stable manner. Low values for $\beta$ increase the time it takes the system to converge and keep the system further away from the target channel occupation. On the other hand, high values for $\beta$ cause the system to \textcolor{black}{oscillate before converging, increasing the potential for situations where it does not converge (equation \ref{equation:limOrig}) or converges to a limit cycle (equation \ref{equation:limSigno})}. High values for $\alpha$ improve convergence speed, but the system converges to a point further away from the target occupation. Therefore, LIMERIC authors explore different values for the algorithm parameters in updates to their original proposal (e.g., \cite{LimericStability} \cite{LimericForwarding}). 

In \cite{LimericStability}, the authors compare DCC mechanisms in dense scenarios, such as winding highways and roads with bridges. They update the parameters for the LIMERIC algorithm as shown in Table \ref{table:limStaPar}. The work in \cite{LimericStability} uses a higher value for the CBR target parameter, \textcolor{black}{which, in combination with the values for $\alpha$ and $\beta$, results in a large variation in the CBR levels to which the algorithm converges at different vehicle densities ($K$ in equation \ref{equation:cbrlimeric})}.

\begin{table}[h]
	\centering
	\caption{Parameter values for LIMERIC from \cite{LimericStability} and \cite{LimericForwarding} }
	\begin{tabular}{|c | c | c |}
		\hline
		\textbf{Parameter} & \textbf{Value in \cite{LimericStability}} & \textbf{Value in \cite{LimericForwarding}}\\[0.5ex]
		\hline
		$\alpha$ & 0.1 & 0.01  \\
		$\beta$ & 0.00167 & 0.001 \\
		CBR Target & 0.79 & 0.65 \\[1ex]
		\hline
	\end{tabular}
	
	\vspace{1ex}
	\small
	{\raggedright \textcolor{black}{\textit{Note:} The value for $\beta$ in \cite{LimericStability} is converted to the equivalent one presented here to account  for the fact that equation 1 in \cite{LimericStability} uses rates instead of $\delta$s and percentages instead of fractions of CBR.}\par}
	
	\label{table:limStaPar}
\end{table}

Another update to LIMERIC is proposed in \cite{LimericForwarding}, where the authors take values for $\alpha$ and $\beta$ that are closer to those adopted by the ETSI Adaptive Approach. It is worth noting that for both updates ---\cite{LimericStability} and \cite{LimericForwarding} --- while the values for these parameters change, the way in which the transmission rate for each \textcolor{black}{vehicle} is updated does not change from the one proposed originally in equation \ref{equation:limOrig}.

\paragraph{ETSI Adaptive Approach}
The Adaptive Approach was defined initially in ETSI TS 102 687 V1.2.1 \cite{etsiNewDcc} in 2018. Its logic is based on LIMERIC, but adds more parameters to control the update rate and to control the minimum and maximum allowed fraction of medium usage to prevent starvation. This algorithm relies on local CBR measurements that are smoothed, and it also permits the use of global CBR measurements when information sharing is supported. The parameters for the standardized mechanism are shown in Table \ref{table:StdPar}.

\begin{table}[h!]
	\centering
	\caption{Parameter values for ETSI Adaptive Approach from \cite{etsiNewDcc}}
	\begin{tabular}{|r | l | p{4.5cm}|}
		\hline
		\textbf{Parameter} & \textbf{Value} & \textbf{Description} \\[0.5ex]
		\hline
		$\alpha$ & 0.016 & Weight given to $\delta(n-1)$ \\
		$\beta$ & 0.0012 & Algorithm parameter \\
		$CBR_{target}$ & 0.68 & The adaptive approach updates $\delta$ so that CBR adapts to this target\\
		$\delta_{max}$ & 0.03 & Upper bound on allowed fraction of medium usage\\
		$\delta_{min}$ & 0.0006 & Lower bound on allowed fraction of medium usage, to prevent starvation under high CBR\\
		$G^+_{max}$ & 0.0005 & Algorithm parameter\\
		$G^-_{max}$ & -0.00025 & Algorithm parameter\\
		$T_{CBR}$ & 100 ms & Interval over which CBR is measured. $\delta$ is updated at twice this interval\\[1ex]
		\hline
	\end{tabular}
	\label{table:StdPar}
\end{table} 

The algorithm that each vehicle executes to update its transmission rate for instant ($n$) is as follows:
\begin{enumerate}
	\item[Step 1:] \begin{equation}  \begin{split}
	CBR_{\textcolor{black}{vehicle}}(n) = 0.5 \cdot CBR_{\textcolor{black}{vehicle}}(n-1) + \\0.5\cdot((CBR_{L} + CBR_{LPrev})/2)
	\end{split} \end{equation}
	\begin{enumerate}
		\item[NOTE 1:] $CBR_{L}$ and $CBR_{LPrev}$ are the last two measurements of $CBR$. $CBR$ is measured at two times the speed at which $\delta$ is calculated. If information sharing is supported, the local measurements are substituted by global measurements (i.e., $CBR_{L}$ is substituted by $CBR_{G}$). 
		
	\end{enumerate}
	\item[Step 2:]    \begin{equation} \begin{split}
	 & \text{If } sign(CBR_{target} - CBR_{\textcolor{black}{vehicle}}(n)) \text{ is positive then } \\
      & \hspace{0.2cm} \delta_{offset} = \\
      & \hspace{0.5cm} min( \beta\cdot(CBR_{target} - CBR_{\textcolor{black}{vehicle}}(n)),  G^+_{max} );  \\
	& \text{Else }  \\
	&\hspace{0.2cm} \delta_{offset} = \\
      & \hspace{0.5cm} max( \beta\cdot(CBR_{target} - CBR_{\textcolor{black}{vehicle}}(n),  G^-_{max}) );  
	\end{split}
       \end{equation}
	\item[Step 3:] \begin{equation}\label{Equation:etsidelta} \begin{split}
	\delta(n) = (1 - \alpha)\cdot\delta(n-1) + \delta_{offset}
	\end{split} \end{equation}
	\item[Step 4:] \begin{equation} \begin{split}
	\text{If }\delta(n) > \delta_{max}, \delta(n) = \delta_{max}
	\end{split} \end{equation}
	\item[Step 5:] \begin{equation} \begin{split}
	\text{If }\delta(n) < \delta_{min}, \delta(n) = \delta_{min}
	\end{split} \end{equation}
\end{enumerate}

The obtained $\delta(n)$ is then used to calculate the time for the next transmission. The value for the transmit duration of the last packet ($t_{on_{pp}}$) is used as a reference to calculate the time to wait for the next transmission ($t_{go}$) to occur after the last transmission ended ($t_{pg}$). This waiting time ($t_{go}$) is constrained to maximum and minimum values established by ETSI EN 302 571 V2.1.1\cite{etsiReqs}, which limits a \textcolor{black}{vehicle} to transmit at intervals no shorter than 25 milliseconds and no longer than 1 second. This is reflected in equation \ref{equation:tgo}.

\begin{equation} \label{equation:tgo} \begin{split}
t_{go} = t_{pg}+min(max(\frac{t_{on_{pp}}}{\delta},25ms),1s)
\end{split} \end{equation}

While the parameter values for ETSI Adaptive Approach differ greatly from the early iterations of LIMERIC, the logic behind them is justified. The value for $\alpha$ is almost one order of magnitude lower, which allows for the aggregate $CBR$ to be closer to $CBR_{target}$ when the system reaches convergence. A problem arises with this low value, because the speed of convergence is slower \cite{Letter}. \textcolor{black}{This also has an effect on fairness in transitory situations}, since the term $(1-\alpha)\cdot\delta$ in equation \ref{Equation:etsidelta} represents the portion of the previous value of $\delta$ that the \textcolor{black}{vehicle} yields for the next calculation. 

On the other hand, $\beta$ is small to account for scenarios with high density of \textcolor{black}{vehicle}s. It is worth noting that the algorithm does not allow $\delta$ to go below $\delta_{min}$, which means the system can handle up to 1133.33 \textcolor{black}{vehicle}s before going over the target occupation.

\paragraph{Dual-$\alpha$ DCC}
While the parameters of the ETSI Adaptive Approach are useful for high and low density scenarios, \textcolor{black}{there is a trade-off when it comes to the speed of convergence metric}. As the work in \cite{Letter} describes, a small value of $\alpha$, as the one used in the ETSI DCC Adaptive Approach, results in a utilization of the medium in steady state situations close to the intended target, but reduces the ability to lower the allocated share of the medium to \mbox{\textcolor{black}{vehicle}s} when needed. This may create periods of congestion and/or unfairness (different \textcolor{black}{vehicle}s in the same situation are allowed to send at different rates, because the algorithm takes time in reducing the allocated share of the medium to some \textcolor{black}{vehicle}s to distribute resources fairly). On the other hand, a large $\alpha$ value results in a utilization of the medium in steady state situations that is farther below the intended target, but with better ability to reduce the allocated share of the medium to \textcolor{black}{vehicle}s when needed.

\textcolor{black}{To address this trade-off, reference \cite{Letter} proposes using a large $\alpha$ value when $\delta$ is decreasing, to help shorten the time it takes the system to converge; while using the standardized value for $\alpha$ the rest of the time, to keep the CBR close to the target utilization.} In particular, the Dual-$\alpha$ algorithm calculates $\alpha$ according to equation \ref{eq:newalpha}:

\begin{equation}
\label{eq:newalpha}
\alpha(n) =
\begin{cases}
\alpha_{high} & \text{if }  
(\delta(n-1) - \delta_{\alpha_{low}}(n)) > th  \\ 
\alpha_{low} & otherwise
\end{cases}
\end{equation}

\begin{table}[tb]
	\centering
	\caption{Parameter values of the Dual-$\alpha$ DCC algorithm from \cite{Letter}}
	\label{tbl:parameters_modified_dcc}
	\begin{tabular}{|c|c|}
		\hline
		\textbf{Parameter}  & \textbf{Value} \\
		\hline
		$\alpha_{low}$ & 0.016 \\
		$\alpha_{high}$ & 0.1 \\   
		$th$ & 0.00001 \\
		\hline
	\end{tabular}
\end{table} 

where $\alpha(n)$ is the value to be used in Step 3 (equation \ref{Equation:etsidelta}) for $\delta$ at time $n$, $\delta(n-1)$ is the previous $\delta$ calculated at $n-1$, $\delta_{\alpha_{low}}(n)$ is a $\delta$ calculated using $\alpha=\alpha_{low}$, and $th$ is a threshold with an heuristic value that improves stability, triggering the change to use $\alpha_{high}$ only when the drop in  $\delta$ is significant enough. The parameters proposed for the Dual-$\alpha$ algorithm are presented in Table \ref{tbl:parameters_modified_dcc}, with the rest of parameters equal to those in Table \ref{table:StdPar}. The results in \cite{Letter} show that the use of two values for $\alpha$ increases the speed of convergence and improves fairness in transitory situations. However, the work in \cite{Letter} is based on analytical results and simulation of transitory scenarios with static \mbox{\textcolor{black}{vehicle}s}. \textcolor{black}{An open question about Dual-$\alpha$ is whether it can achieve the performance of the ETSI DCC Adaptive Approach in steady state scenarios.}

\section{Simulation Model Description}
\label{sec:simdesc}
\subsection{Simulation Scenarios}
We evaluate the performance of the ETSI Adaptive Approach DCC mechanism \cite{etsiNewDcc} and compare it with two related algorithms: LIMERIC from \cite{LimericStability}, and Dual-$\alpha$ \cite{Letter} \textcolor{black}{(the parameters of the three algorithms have been described in previous section)}. LIMERIC from \cite{LimericStability} was chosen because of its CBR target (0.79), which is different to those in \cite{LimericOriginal} and \cite{LimericForwarding}, with the latter having parameters that are close to those in the ETSI standard. The logic behind the 0.79 target is to compensate for a high $\alpha$ parameter, which keeps the convergence point away from the target. Henceforth, LIMERIC from \cite{LimericStability} will be referred to as LIMERIC 0.79. We also compare these algorithms with a baseline benchmark obtained in absence of a DCC mechanism and relying only on the EDCA mechanism to control transmissions.

The algorithms are evaluated in three dynamic scenarios: one where the algorithms are in a steady state situation and two that test their reaction in transitory situations. The first test scenario, called Oval Scenario, is a 7.75 km oval road with four lanes --- two in each direction --- that has both long straight stretches and curves at the edges. The aim of this scenario is to test the mechanisms in situations where vehicle dynamics stay in a stationary state regarding average speed and acceleration. Several vehicle densities are tested, ranging from those where traffic is in free flow to those where average speeds drop to 34\% of the maximum speed set by the road. 

The first transitory scenario is a road with a traffic light that controls the entrance into a freeway, typical of major motorways beginning in or crossing urban areas, such as interstate roads I-35 and I-78 in the United States. In this scenario, speed of convergence and its effect are measured. A group of cars wait in the red light before a curve that will head them into the freeway. While they are waiting in the red light, their dynamic CAM message generation rate drops to a minimum. When the group starts moving, vehicle dynamics change, and the message rate allowance has to adapt rapidly and keep channel usage within boundaries.

The last scenario, called Junction Scenario, is an intersection where a small group of 25 cars in a two-lane motorway incorporates into a larger group of 300 vehicles circulating in a six-lane highway, in which fairness is evaluated. Cars in the small group sense a lower channel occupation, which allow them to transmit most of the messages they generate, while the larger group senses a higher occupation which causes them to modulate their transmission rate accordingly. The objective of this scenario is to evaluate how fast the two groups adjust their transmission and generation rates after merging.

\begin{table}[tb]
	\centering
	\caption{Simulation parameters}
	\label{tbl:simpars}
	\begin{tabular}{| l | l |}
		\hline
		\textbf{Parameter}  & \textbf{Values} \\
		\hline
		Access layer protocol & ITS-G5 (IEEE 802.11p) \\
		Data rate & 6 Mbit/s \\
		Transmit power & 126 mW \\
		Channel bandwidth & 10 MHz at 5.9 GHz \\
		Path loss model & Two-Ray Interference Model \\
		Maximum sensing rage & 750 m \\
		\hline
	\end{tabular}
\end{table}

Simulations are performed in Artery \cite{Artery}, a simulation toolkit that implements ETSI ITS-G5. It runs on OMNET++ and it is based on Veins \cite{Veins}, which provides the model for the physical layer, in this case, the Two-Ray interference model\cite{TwoRay}. Parameters for Artery can be seen in \mbox{Table \ref{tbl:simpars}}. Artery uses SUMO (Simulation of Urban MObility) to simulate traffic. SUMO models individual vehicles and their interactions using models for car-following, lane-changing and intersection behavior. For our simulations, the default attributes for these models were left as is, with flows consisting of only the default vehicle type (a passenger car). Statistics are collected from at least 25 randomly-selected \textcolor{black}{vehicle}s. In the cases where average values are shown, 95\% confidence intervals are calculated.

\subsection{Performance metrics}
\paragraph{Channel Busy Ratio and $\delta$}
The final goal of the DCC mechanisms we analyze is to keep the CBR below a target, so one of the metrics we use to analyze the performance of these DCC mechanisms is how CBR, and values of $\delta$ that define it, behave in the simulated scenarios. However, these are not the only metrics we analyze. 

Diverse types of applications require different types of metrics. There are services that rely on messages being delivered to as many \textcolor{black}{vehicles} as possible (e.g., emergency brake lights, rescue vehicles approaching, etc.), while there are services where the information should be updated with a certain frequency in order to be reliable (e.g., Cooperative Awareness, slow or stationary vehicles, road conditions, among others). That is why the following metrics were chosen.

\paragraph{Packet Delivery Ratio}
Packet Delivery Ratio (PDR) is the percentage of receptions registered in remote \textcolor{black}{vehicle}s compared to all the messages a host \textcolor{black}{vehicle} generates. PDR can be affected by factors such as the distance between transmitting and receiving \textcolor{black}{vehicles} (due to path loss and interference), or in certain cases due to collisions (e.g., because of the hidden node problem).

\paragraph{Inter-packet gap}
Inter-packet gap is the time between two messages received from the same \textcolor{black}{vehicle}. While losses can affect how frequently packets from a neighbor are received, inter-packet gap is also influenced by the frequency messages are generated (e.g., when the cross-layer DCC mechanism restricts message generation). This has an effect on the performance of awareness services, where lack of updated information could lead to a difference between the expected and actual location of a remote \textcolor{black}{vehicle}. \textcolor{black}{This work presents the results of the 95th percentile inter-packet gaps to better complement the information provided by the PDR parameter.}

\section{Simulation Results}
\label{sec:simres}
\begin{figure}[t]
	\centering
	\includegraphics[width=245px,height=200px,clip,keepaspectratio]{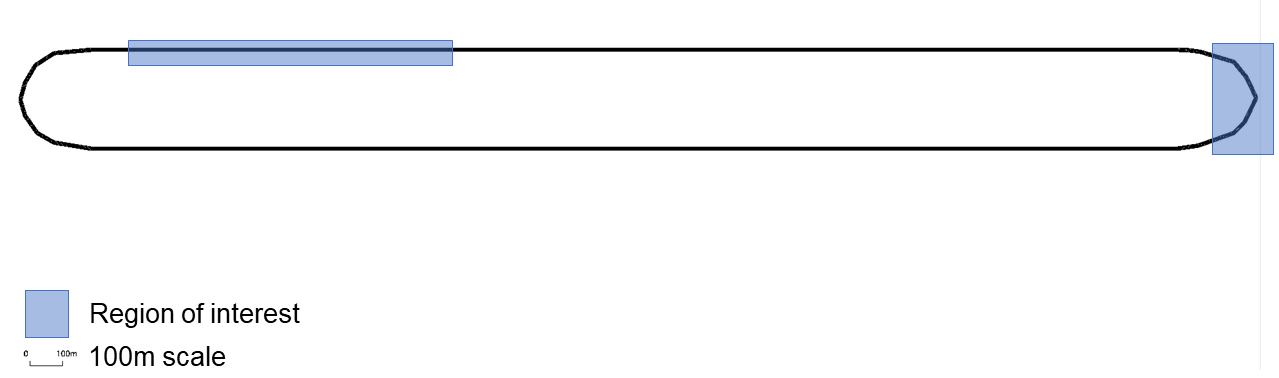}
	\caption{Oval Scenario}
	\label{fig:ovalscenario}
\end{figure}

\paragraph{Oval Scenario}
Fig. \ref{fig:ovalscenario} shows the schematic for this scenario, highlighting the Regions of Interest (ROI): a straight stretch and a curve. Data about sent messages is collected from cars passing through these areas, and receptions are recorded in \textcolor{black}{vehicle}s receiving these messages in the whole scenario, whether they are in the ROI or not. The simulation runs for a period of time until the target density is reached. Then, measurements are taken in the two ROIs during 120 s. Confidence intervals are shown in the figures only when they differ significantly from the average values.

Six different vehicle densities are used, from 10 to 60 vehicles/km per lane. Vehicle speeds are limited by the maximum speed limit established by the road (33 m/s), and by the vehicle density, with higher densities causing lower average speeds, as shown in Table \ref{tbl:ovalspeeds}. Higher densities do not only reduce average speeds, but also increase other movements such as braking, accelerating, and steering.

For this scenario, initially, we only consider the CAM service (DP2 messages), whose generation follows the rules established in \cite{etsiCAM}, where changes in the vehicle dynamics such as speed and heading, along with DCC Facilities, trigger message generation within lower and upper boundaries of 0.1 and 1 seconds respectively. Between these two limits, message generation rate is controlled by feedback information coming from the DCC at lower layers through the Management entity (Fig. \ref{fig:dccarchitecture}), which provides the minimum time interval between two consecutive generations. If the time elapsed since the last CAM generation exceeds \textcolor{black}{the time interval provided} by DCC, the triggering conditions (i.e., vehicle dynamics) are checked in order to generate a message.

\begin{table}[tb]
	\centering
	\caption{Average speed for different densities}
	\label{tbl:ovalspeeds}
	\begin{tabular}{|c|c|}
		\hline
		\textbf{Density (veh/km per lane)}  & \textbf{Average speed (m/s)} \\
		\hline
		10 & 26.50 \\
		20 & 22.33 \\
		30 & 16.75 \\
		40 & 14.20 \\
		50 & 12.45 \\
		60 & 11.40 \\
		\hline
	\end{tabular}
\end{table} 

\begin{figure}[h]
	\centering
	\includegraphics[width=205px,clip,keepaspectratio]{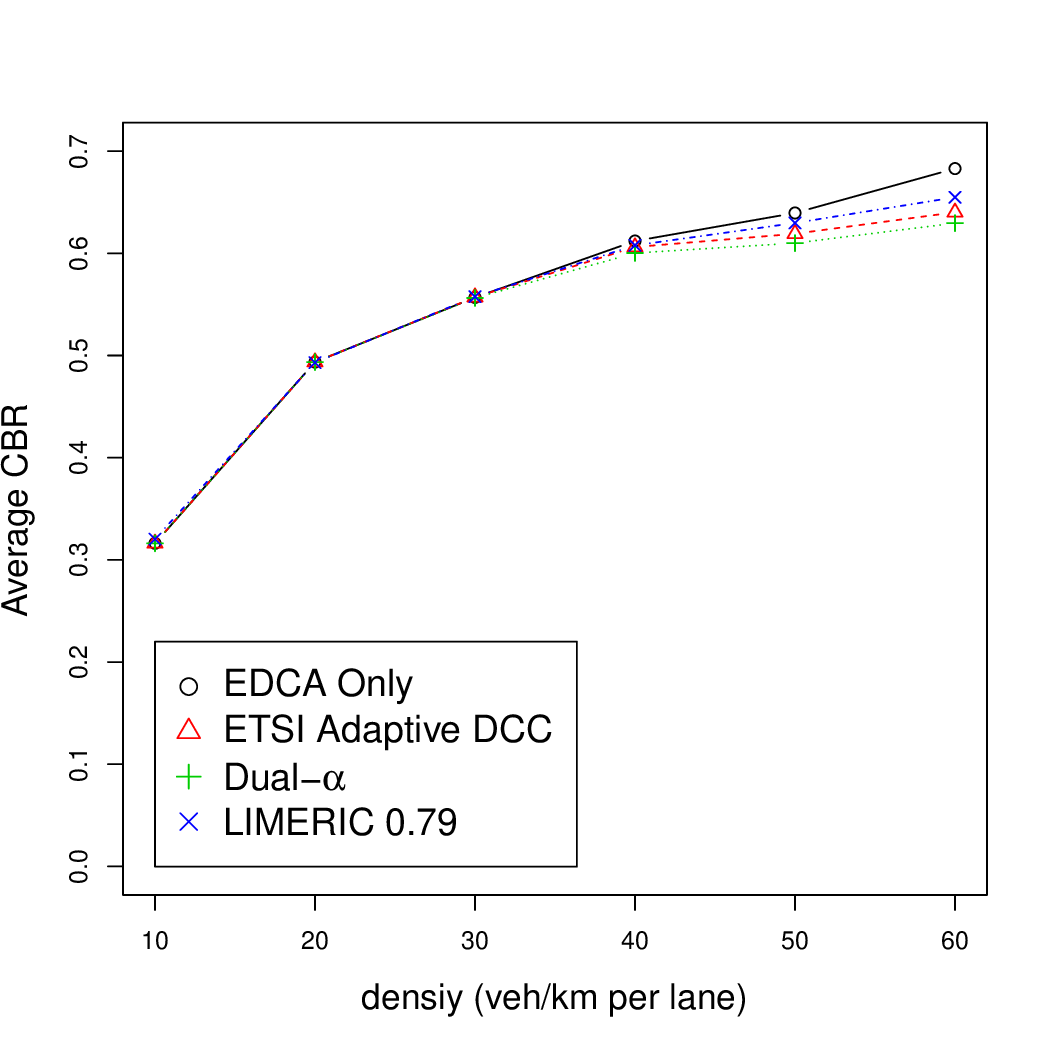}
	\caption{Average channel occupation for densities of 10-60 veh/km per lane}
	\label{fig:avgcbr}
\end{figure}

\begin{table}[tb]
	\small
	\centering
	\caption{Average time between the generation of consecutive CAM messages}
	\label{tbl:avggen}
	\textbf{Curve}
	\begin{tabular}{|c|c|c|c|c|}
		\hline
		\textbf{Density}  & \textbf{EDCA only}  & \textbf{ETSI}  & \textbf{Dual-$\alpha$}  & \textbf{LIMERIC 0.79} \\
		\hline
		10 & 0.1591 s & 0.1591 s & 0.1591 s & 0.1591 s\\
		20 & 0.1612 s & 0.1612 s & 0.1612 s & 0.1612 s\\
		30 & 0.2316 s & 0.2316 s & 0.2316 s & 0.2316 s \\
		40 & 0.2564 s & 0.2564 s & 0.2570 s & 0.2588 s\\
		50 & 0.2957 s & 0.3088 s & 0.3087 s & 0.2990 s\\
		60 & 0.2773 s & 0.2781 s & 0.2778 s & 0.2779 s\\
		\hline
	\end{tabular}

	\textbf{Straight Stretch} \\
	\begin{tabular}{|c|c|c|c|c|}
		\hline
		\textbf{Density}  & \textbf{EDCA only}  & \textbf{ETSI}  & \textbf{Dual-$\alpha$}  & \textbf{LIMERIC 0.79} \\
		\hline
		10 & 0.1642 s & 0.1642 s & 0.1642 s & 0.1642 s\\
		20 & 0.2224 s & 0.2224 s & 0.2224 s & 0.2224 s\\
		30 & 0.2153 s & 0.2153 s & 0.2153 s & 0.2153 s\\
		40 & 0.2745 s & 0.2745 s & 0.2746 s & 0.2855 s\\
		50 & 0.3057 s & 0.3292 s & 0.3322 s & 0.3179 s\\
		60 & 0.3069 s & 0.3493 s & 0.3523 s & 0.3228 s\\
		\hline
	\end{tabular}
\end{table}

\begin{figure}[h]
	\centering
	\includegraphics[width=205px,clip,keepaspectratio]{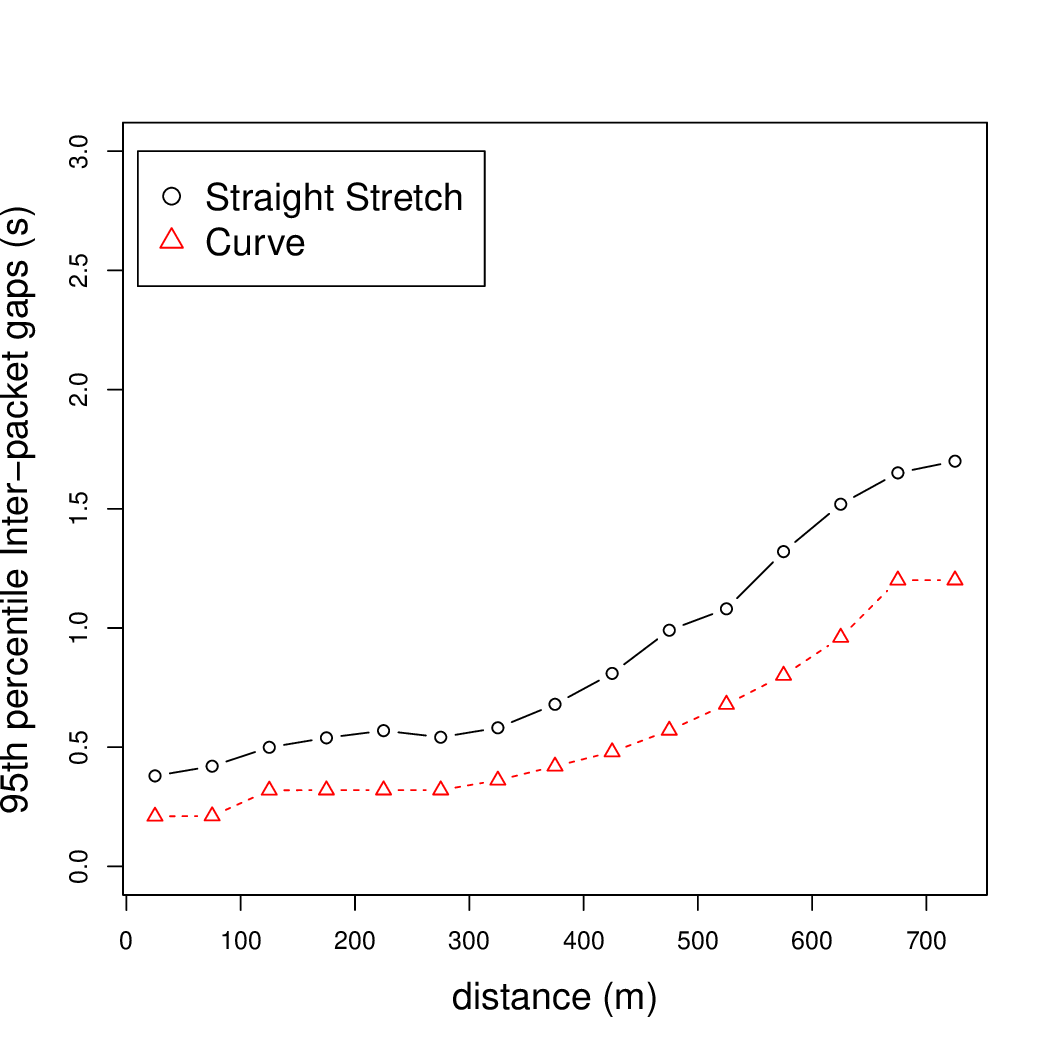}
	\caption{\textcolor{black}{95th percentile inter-packet gaps for a density of 20 veh/km per lane}}
	\label{fig:gapsd20}
\end{figure}

\begin{figure}[h]
	\centering
	\includegraphics[width=205px,clip,keepaspectratio]{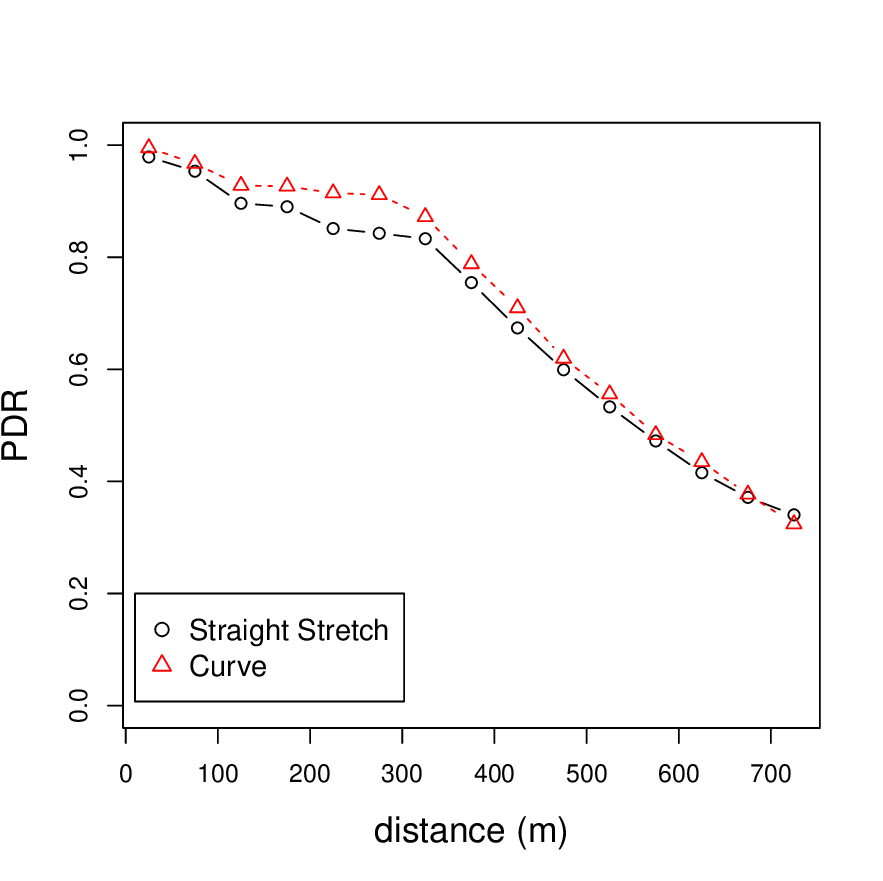}
	\caption{Average Packet Delivery Ratio for a density of 20 veh/km per lane}
	\label{fig:pdrd20}
\end{figure}

Fig. \ref{fig:avgcbr} shows that the channel occupancy stays well below the target value of the ETSI Adaptive Approach and \mbox{Dual-$\alpha$} (0.68), and that of LIMERIC 0.79 (0.79) for densities up to 40 veh/km per lane. This means that DCC allows all the traffic to be handled by EDCA, which reflects on the metrics, where there is no difference in the performance of the different DCC algorithms at lower densities, as they do not restrict message generation. 

The average time between the generation of two CAM messages, for both ROIs, are shown in Table \ref{tbl:avggen}, which also includes a column (EDCA only) where feedback from DCC is not considered at the CA service, and DCC does not \textcolor{black}{limit message rates} in the Access Layer. For all densities, due to the feedback from DCC Facilities to CA service, CAM messages are generated, at most, at the rate allowed by the DCC mechanism. Therefore, in absence of another type of traffic in a \textcolor{black}{vehicle}, the generation rate is equal to the transmission rate.

Fig. \ref{fig:gapsd20} represents the 95th percentile inter-packet gaps separated for each ROI for a density of 20 vehicles/km per lane. For distances above 300 meters, the inter-packet gaps increase significantly, since losses are caused by the distance between \textcolor{black}{vehicles} and collisions mainly due to hidden nodes, which is also reflected by the values for PDR in Fig. \ref{fig:pdrd20}. Since DCC mechanisms do not exercise influence, results are the same for all algorithms. \textcolor{black}{We performed experiments for the two ROIs (curve and straight stretch) and similar results were obtained, with the straight stretch being the most interesting since it exhibits a higher load due to the scenario topology. For this reason, from this point forward, we show results only for the straight stretch ROI}.

\begin{figure}[h]
	\centering
	\includegraphics[width=205px,clip,keepaspectratio]{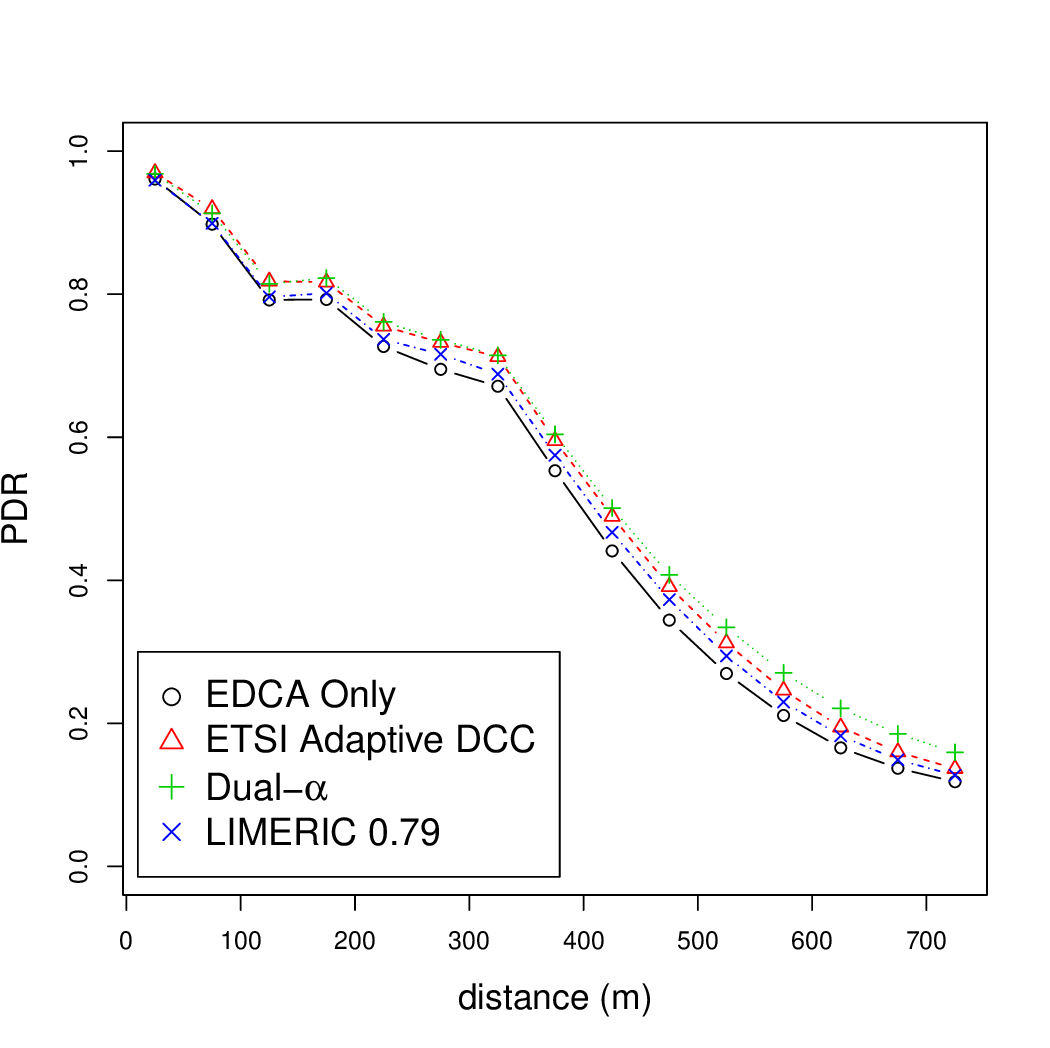}
	\caption{Average Packet Delivery Ratio for a density of 50 veh/km per lane (straight stretch)}
	\label{fig:pdrd50r}
\end{figure}

\begin{figure}[h]
	\centering
	\includegraphics[width=205px,clip,keepaspectratio]{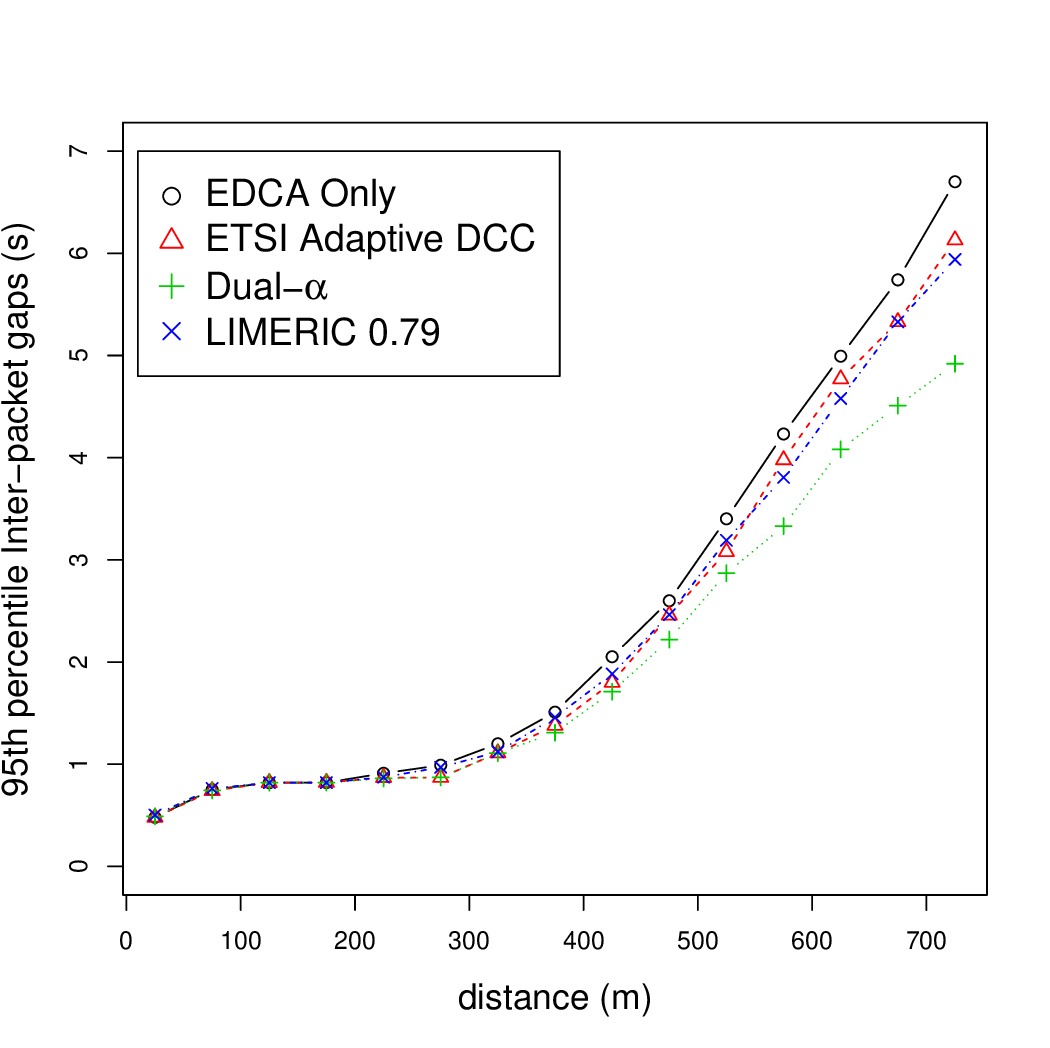}
	\caption{\textcolor{black}{95th percentile inter-packet gaps for a density of 50 veh/km per lane (straight stretch)}}
	\label{fig:gapsd50r}
\end{figure}

A gap between average CBR values appears when the density increases to 50 and 60 vehicles/km per lane (Fig. \ref{fig:avgcbr}). \textcolor{black}{The DCC mechanisms start acting so that channel occupation stays below their targets. There are small differences in occupation, which are due to the parameters for each algorithm: at these high densities, the target CBR set for LIMERIC 0.79, results in larger average utilization}. Table \ref{tbl:avggen} shows how, for 50 and 60 vehicles/km per lane densities, DCC mechanisms delay the generation of CAM messages longer than what dynamics demand (as seen in the EDCA column). 

There is an effect on PDR values, which start dropping below 80\% success rates at shorter distances as shown in Fig. \ref{fig:pdrd50r} for a density of 50 vehicles/km per lane. Furthermore, \textcolor{black}{95th percentile inter-packet gaps are similar for distances less than 300 m, and at longer ranges they stay consistently lower when using DCC mechanisms than when using only EDCA, as shown in Fig. \ref{fig:gapsd50r}. Finally, {Dual-$\alpha$} seems to be able to capitalize on its ability to adapt to changes to achieve lower inter-packet gaps.}

These results, in scenarios that comply with standardized rules established in \cite{etsiCAM} and \cite{etsiNewDcc}, show the positive effect of DCC mechanisms. All DCC algorithms generate and send less messages but, since they use the medium more efficiently by not congesting it, they perform better in the PDR and inter-packet gap metrics \textcolor{black}{than using only EDCA}. 

However, it is not only under conditions where many \mbox{\textcolor{black}{vehicle}s} are within range that channel occupation becomes a concern. In real-life scenarios, multiple services and applications may coexist, sharing the medium and adding to the channel occupation. 

In order to evaluate further the performance of the different DCC mechanisms, we select two of the densities --- 20 and 50 vehicles/km per lane --- and \textcolor{black}{perform a test where another type of traffic, background traffic, coexists with the CAM messages used in the scenarios shown above. The background traffic is generated at the facilities layer in such a way that \mbox{\textcolor{black}{vehicle}s} always have a packet to transmit, so that the channel occupation will tend to be the maximum allowed by the DCC mechanisms. Background traffic has a data profile with a lower priority (DP3) compared to the CAM traffic (CAM messages are DP2). This means that background messages will be sent when the DCC mechanism allows sending a packet and there is not a CAM message waiting to be sent (i.e., using the capacity of the medium not used by the CAM traffic). This configuration shows the performance of the DCC mechanisms when the offered traffic is larger than the traffic the medium can accept, and the DCC mechanisms have to work constantly to keep the medium utilization at the target value to avoid congestion. We do not present the results for EDCA only, since its performance is much worse in this congested scenario (e.g., for a density of 20 vehicles/km per lane with background traffic, the 95th percentile inter-packet gap for CAM traffic is 4.62 seconds at 400 meters).}  

\begin{figure}[h]
	\centering
	\includegraphics[width=205px,clip,keepaspectratio]{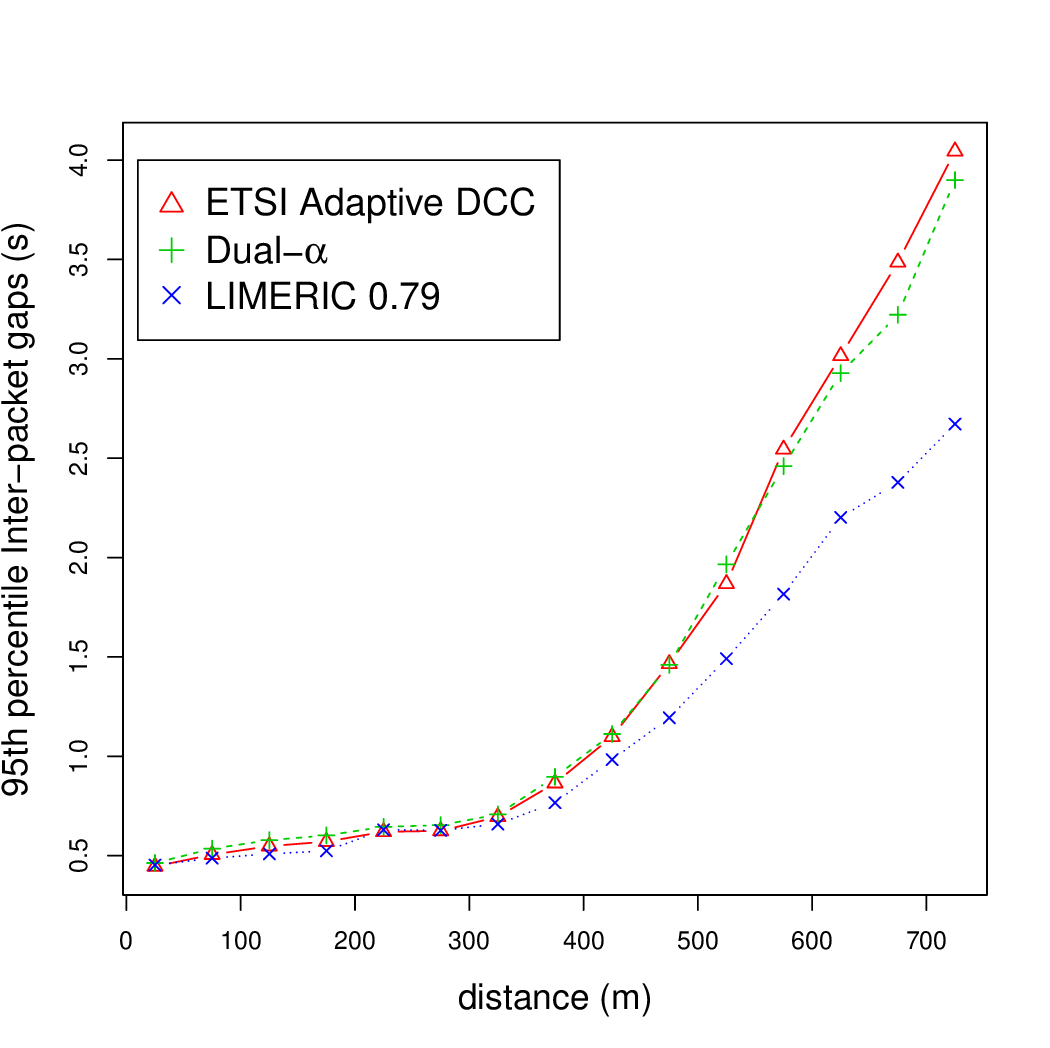}
	\caption{\textcolor{black}{95th percentile inter-packet gaps for DP2 messages in a density of 20 veh/km per lane (straight stretch) with DP2 and DP3 traffic}}
	\label{fig:gapsd20roh}
\end{figure}

\begin{figure}[h]
	\centering
	\includegraphics[width=205px,clip,keepaspectratio]{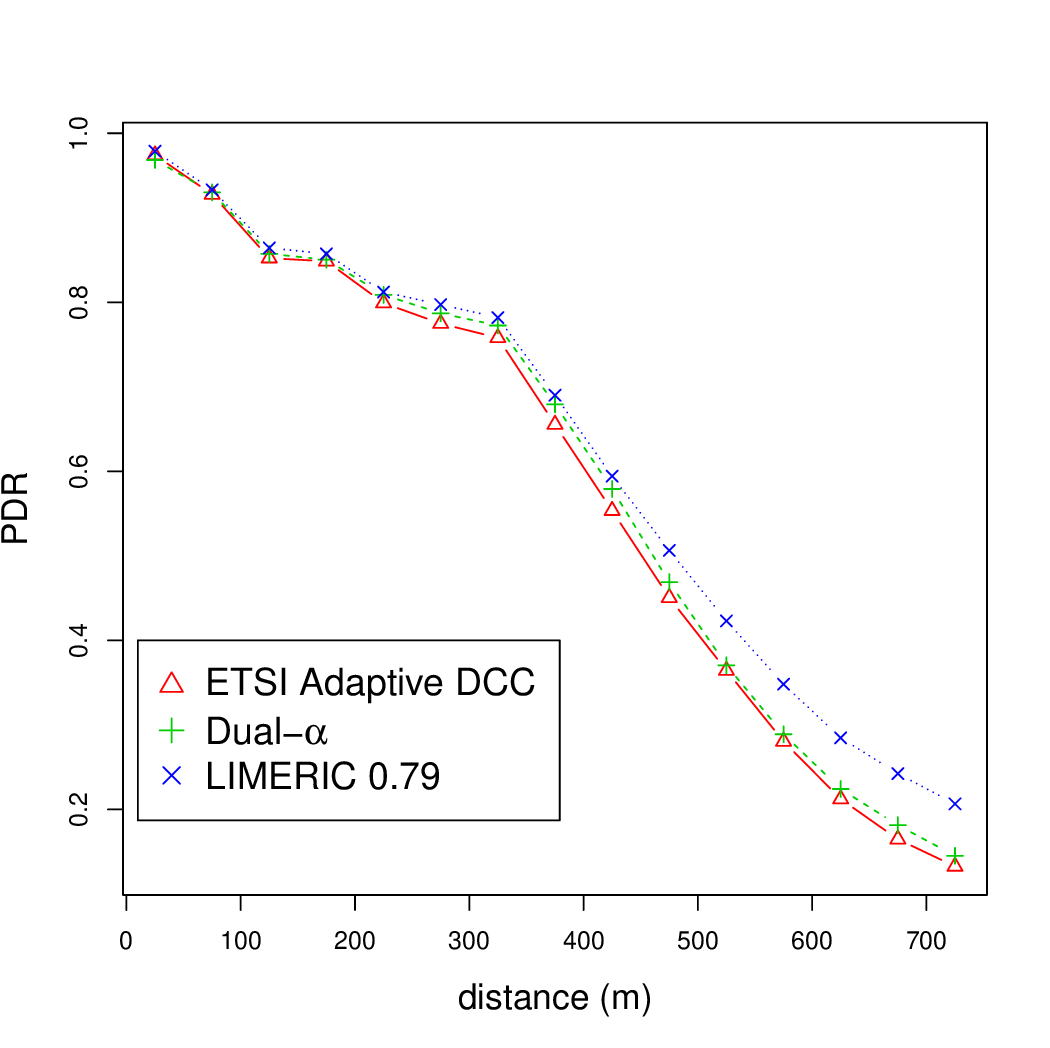}
	\caption{Average Packet Delivery Ratio for DP2 messages in a density of 20 veh/km per lane (straight stretch) with DP2 and DP3 traffic}
	\label{fig:pdrd20roh}
\end{figure}

For the lowest density (20 vehicles/km per lane), Fig.  \ref{fig:gapsd20roh} shows \textcolor{black}{the effect of background traffic on the inter-packet gap metric for CAM messages}. This effect can be noted when comparing \textcolor{black}{this figure} with Fig. \ref{fig:gapsd20}. There is also an effect on PDR, which is noticeable on Fig. \ref{fig:pdrd20roh} when compared with Fig. \ref{fig:pdrd20}. As expected, successful receptions decrease in the presence of background traffic. \textcolor{black}{LIMERIC 0.79 keeps inter-packet gaps lower, in particular at longer distances, when compared to the ETSI and \mbox{Dual-$\alpha$} DCC algorithms. This is due to the fact that, at low densities, LIMERIC 0.79 converges to an average utilization that is lower (0.58 in this experiment) than the ones for the ETSI Adaptive DCC and \mbox{Dual-$\alpha$} mechanisms (0.63 and 0.61 respectively in this experiment). The  downside for LIMERIC 0.79 is that it could send less background traffic successfully, in particular, DP3 throughput with LIMERIC 0.79 is between 38\% and 59\% lower (2.14 and 2.46 msg/s compared to 1.54 msg/s) than the ETSI Adaptive DCC and \mbox{Dual-$\alpha$} approaches respectively.}

\begin{figure}[h]
	\centering
	\includegraphics[width=205px,clip,keepaspectratio]{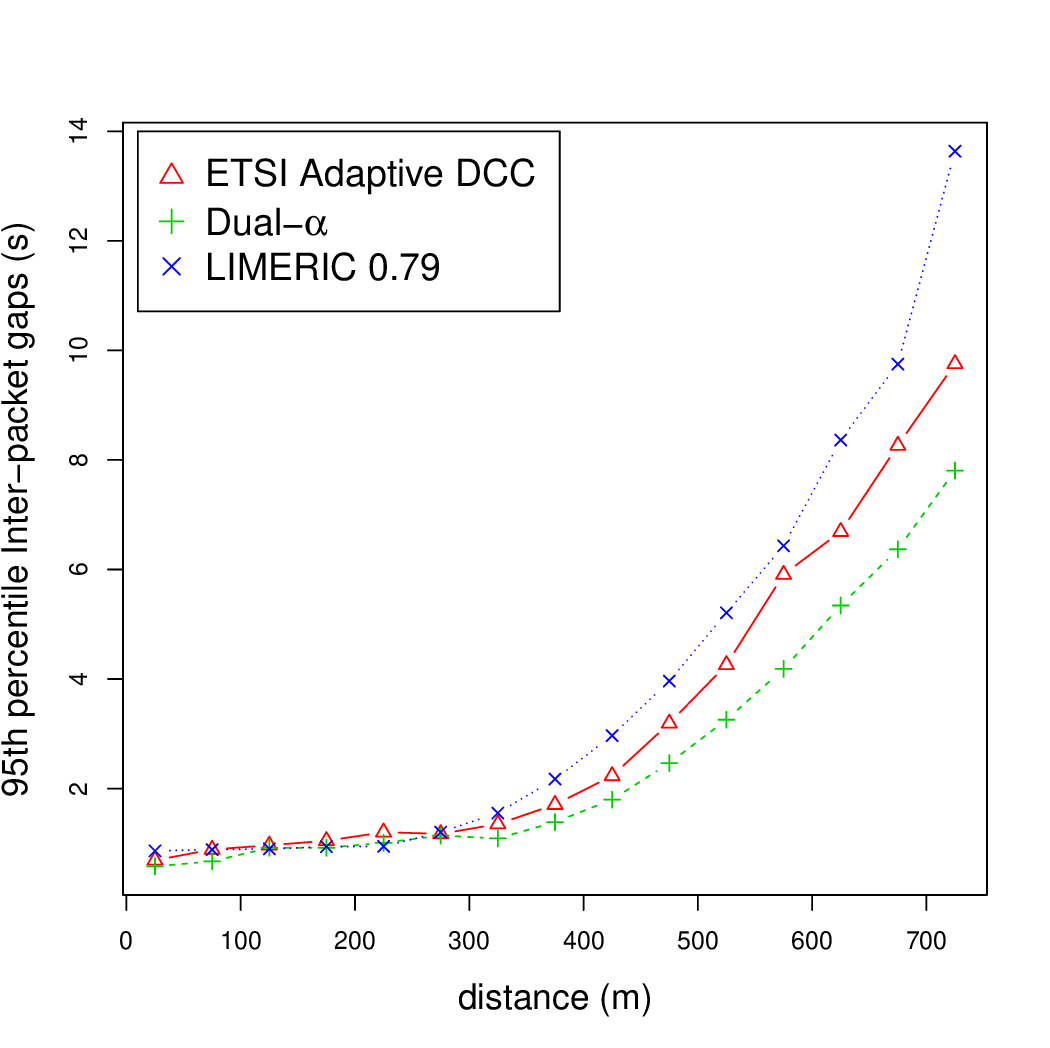}
	\caption{\textcolor{black}{95th percentile inter-packet gaps for DP2 messages in a density of 50 veh/km per lane (straight stretch) with DP2 and DP3 traffic}}
	\label{fig:gapsd50roh}
\end{figure}
\begin{figure}[h]
	\centering
	\includegraphics[width=205px,clip,keepaspectratio]{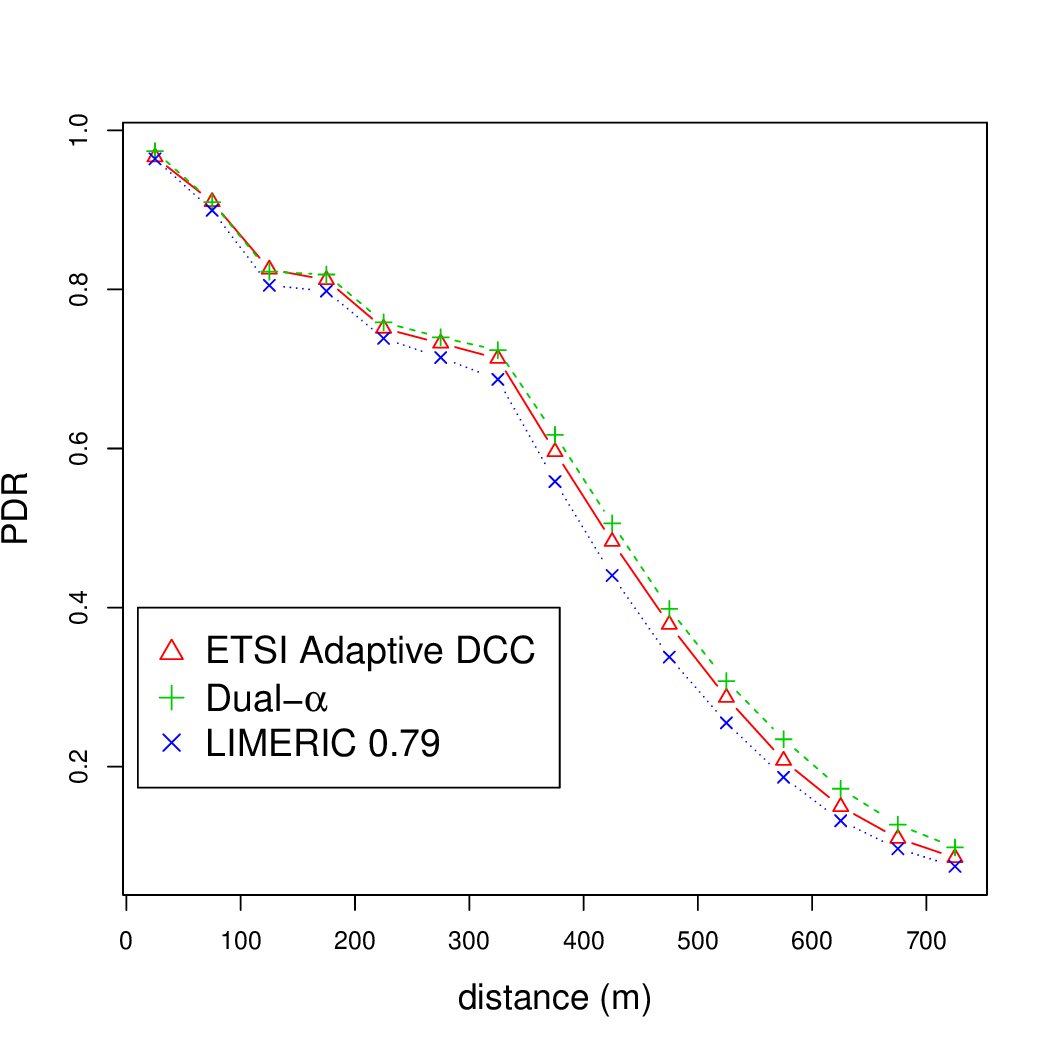}
	\caption{Average Packet Delivery Ratio for DP2 messages in a density of 50 veh/km per lane (straight stretch) with DP2 and DP3 traffic}
	\label{fig:pdrd50roh}
\end{figure}

For the 50 vehicles/km per lane density, Fig. \ref{fig:gapsd50roh} confirms that \mbox{\textcolor{black}{vehicle}s} can keep track of more distant neighbors when DCC mechanisms use stricter parameters (i.e., lower CBR targets). Results for the PDR metrics are shown in Fig. \ref{fig:pdrd50roh}. When comparing these results to those in Fig. \ref{fig:pdrd50r} and \ref{fig:gapsd50r}, it can be noted that mechanisms with stricter parameters manage to \textcolor{black}{perform better} even with background traffic, thanks to the higher priority of the CAM messages, which means that background traffic is only sent when there are available resources. \textcolor{black}{On the other hand, LIMERIC 0.79 pays a price in performance for having a higher CBR target that at high densities translates to a high utilization (an average of 0.684 in this experiment): background traffic is sent because the algorithm considers that there are enough resources on the medium, and its performance (PDR and inter-packet gap) for CAM traffic degrades}.

\begin{table}[tb]
	\small
	\centering
	\caption{Average time between generation of consecutive CAM messages with background traffic}
	\label{tbl:avggenoh}
	\textbf{Straight Stretch}
	\begin{tabular}{|c|c|c|c|c|}
		\hline
		\textbf{Density}  & \textbf{EDCA only}  & \textbf{ETSI}  & \textbf{Dual-$\alpha$}  & \textbf{LIMERIC 0.79} \\
		\hline
		20 & 0.2224 s & 0.2270 s & 0.2414 s & 0.2242 s\\
		50 & 0.3057 & 0.4114 s & 0.3842 s & 0.3416 s\\
		\hline
	\end{tabular}
\end{table}

Table \ref{tbl:avggenoh} shows that higher levels of channel congestion cause a drop in the frequency with which CAM messages are generated, which is an expected result following CAM  generation rules from \cite{etsiCAM}, where the first limit to a CAM generation is the message rate provided by DCC Facilities, which is the rate at which channel conditions allow \mbox{\textcolor{black}{vehicle}s} to transmit. The limit provided by DCC is useful since this feedback allows CAM messages to have a better probability to reach their destination and update their neighbors. When a \textcolor{black}{vehicle} only handles CAM traffic, this mechanism also guarantees that the sent CAM messages contain fresh information, since messages are generated just when it is possible to send them (i.e., the DCC at the access layer will be open).

However, a phenomenon arises when CAM messages coexist with background traffic. If the value of $\delta$ allows \mbox{\textcolor{black}{vehicles}} to transmit CAM messages and some background traffic, the rate provided by DCC Facilities might not coincide with the next time the \mbox{\textcolor{black}{vehicle}} is permitted to transmit (i.e., when DCC allows it). DCC Facilities provides the CA service with a message rate, which is the interval stemming from $\delta$ (i.e., $\frac{t_{on_{pp}}}{\delta}$ in equation \ref{equation:tgo}), while DCC Access updates the time for the next transmission (i.e., $t_{go}$ in equation \ref{equation:tgo}) after any packet is sent. If a background packet is transmitted, $t_{go}$ is updated, producing a time divergence which will cause CAM messages to wait in the DCC queues until their transmission is allowed. This phenomenon causes an increase in the time between a CAM message generation and its reception in a remote \mbox{\textcolor{black}{vehicle}}. 

Table \ref{tbl:avge2ed} shows the values for \textcolor{black}{application layer end-to-end delay for CAM messages (i.e., the period between a CAM generation at the source facilities layer and its reception in a remote \mbox{\textcolor{black}{vehicle}'s} CA service)} in our simulations for the 50 vehicle/km per lane density, with and without background traffic. The end-to-end delay for the CAM messages increases significantly in the scenario with background traffic. If the time that passes between a CAM is generated and its consumption in a remote \textcolor{black}{vehicle} is long enough, it means that its information can be outdated, leading to undesirable effects such as tracking errors. This phenomenon, along with the effects it brings, requires further exploration in a future work.

\begin{table}[tb]
	\centering
	\caption{Average end-to-end delay for CAM messages in the 50 veh/km per lane density}
	\label{tbl:avge2ed}
	\textbf{Straight Stretch}
	\begin{tabular}{|c|c|c|c|}
		\hline
		\textbf{Traffic}  & \textbf{ETSI}  & \textbf{Dual-$\alpha$}  & \textbf{LIMERIC 0.79} \\
		\hline
		Only DP2 & 1.028 ms & 1.010 ms & 1.047 ms\\
		DP2 \& DP3 & 164.88 ms & 152.58 ms & 164.55 ms\\
		\hline
	\end{tabular}
\end{table}

The results of the Oval Scenario, where the performance of DCC mechanisms is measured in a steady state situation, bring several conclusions: 1) congestion control is needed, since it keeps metrics in acceptable levels when compared to the absence of DCC mechanisms starting at medium \mbox{\textcolor{black}{vehicle}s} density levels; 2) \textcolor{black}{The combination of parameters in LIMERIC 0.79 causes underutilization of the medium with a low number of vehicles, and over-utilization in high densities, which affects the algorithm's performance}; and 3) \mbox{Dual-$\alpha$} manages to achieve a performance similar\textcolor{black}{, or even slightly better,} to the ETSI mechanism in steady state situations.

\paragraph{Traffic Light Scenario}
\label{convergence}
Once we have studied the performance of the DCC algorithms in steady state situations, it is important to evaluate how they perform in transitory situations, where speed of convergence is a problem that can arise. This scenario, in which we evaluate convergence, \textcolor{black}{is one where a group of 300 cars distributed in six lanes stop in a traffic light, and the dynamics cause vehicles to generate and transmit CAM messages at a minimum rate (1 Hz)}. This low message rate lowers the channel occupation, which brings $\delta$ to a high value close to $\delta_{max}$. When cars start moving, in this case entering a freeway in the same fashion that cars enter the south end of I-35 in the United States, dynamics bring message generation to a higher rate, close to the maximum of 10 Hz established by \cite{etsiCAM}. 

\begin{figure}[h]
	\centering
	\includegraphics[width=205px,clip,keepaspectratio]{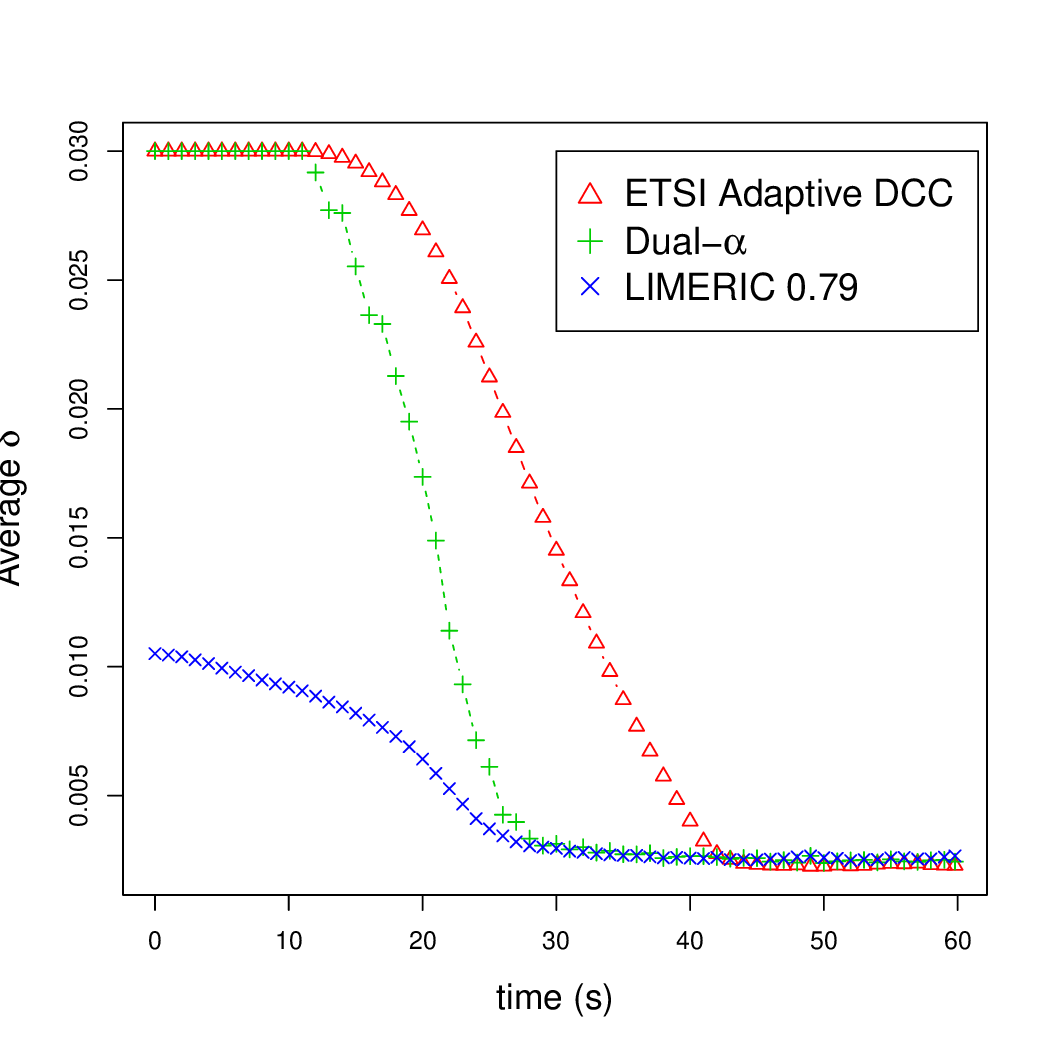}
	\caption{Evolution of $\delta$ for the Traffic Light Scenario}
	\label{fig:deltatl}
\end{figure}

\begin{figure}[h]
	\centering
	\includegraphics[width=205px,clip,keepaspectratio]{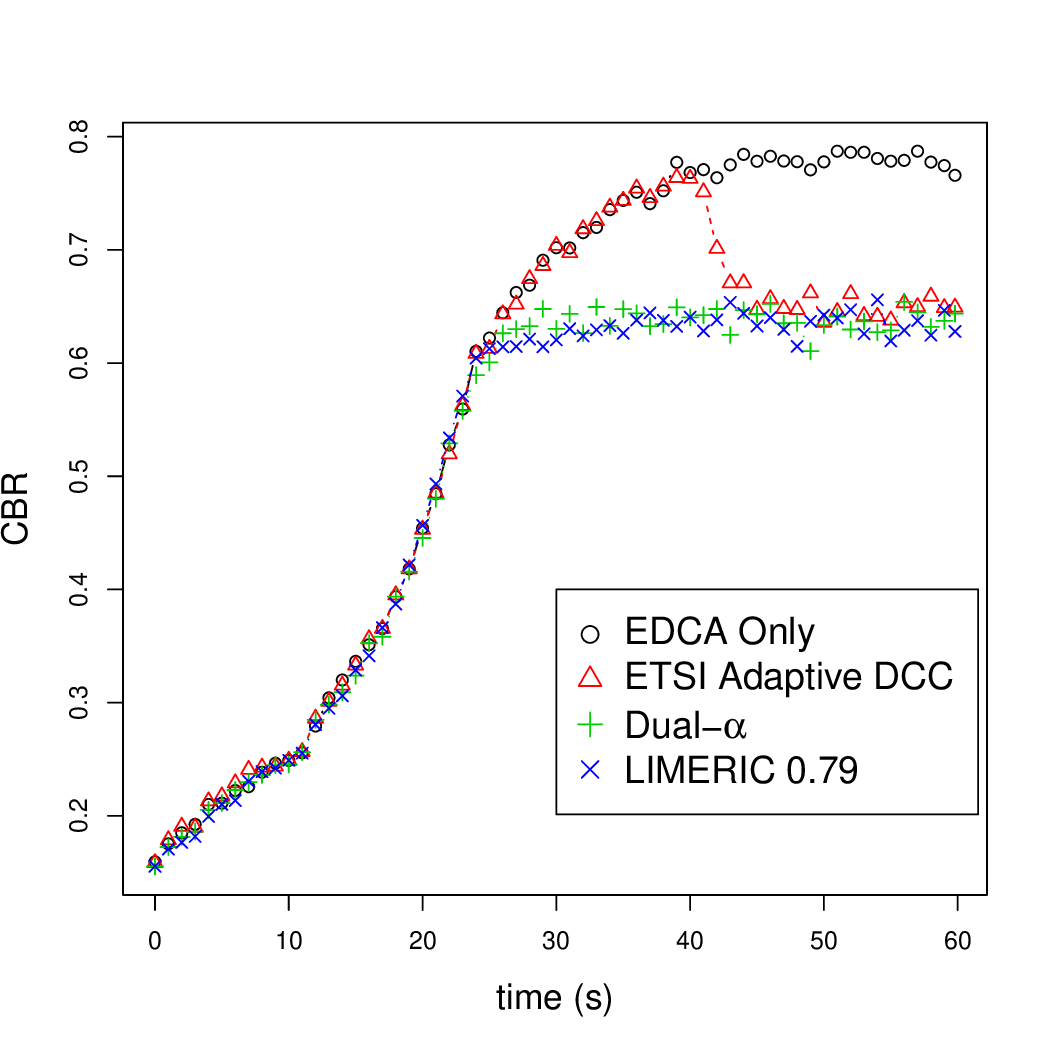}
	\caption{Evolution of channel occupation for the Traffic Light Scenario}
	\label{fig:cbrtl}
\end{figure}

As it can be seen in Fig. \ref{fig:deltatl}, $\delta$ lowers to a value that corresponds to the size and dynamics of the moving group of vehicles after the traffic light allows them to advance ($t=0$). Dual-$\alpha$ reaches this value in half the time it takes the ETSI Adaptive Approach to do so. The effect of \mbox{\textcolor{black}{vehicle}s} having a $\delta$ that does not correspond to the conditions they are currently facing is shown in Fig. \ref{fig:cbrtl}, where there is a stretch of more than ten seconds where the ETSI Adaptive Approach algorithm does not manage to keep CBR below the 0.68 target and it effectively behaves as if there were not a DCC mechanism present (i.e., equal to EDCA only). \textcolor{black}{It can also be noted that, while the value of $\alpha$ in the LIMERIC 0.79 algorithm helps it decrease $\delta$ rapidly, it starts from a value that is lower than those for the ETSI Adaptive Approach and Dual-$\alpha$. This is an example of the underutilization effect that stems from the parameter combination in LIMERIC 0.79 --- even with a low utilization of the medium, vehicles are limited in the traffic they are allowed to send (much lower than the maximum expected, $\delta_{max}$, in the ETSI standard).}

\begin{figure}[h]
	
	\centering
	\includegraphics[width=205px,clip,keepaspectratio]{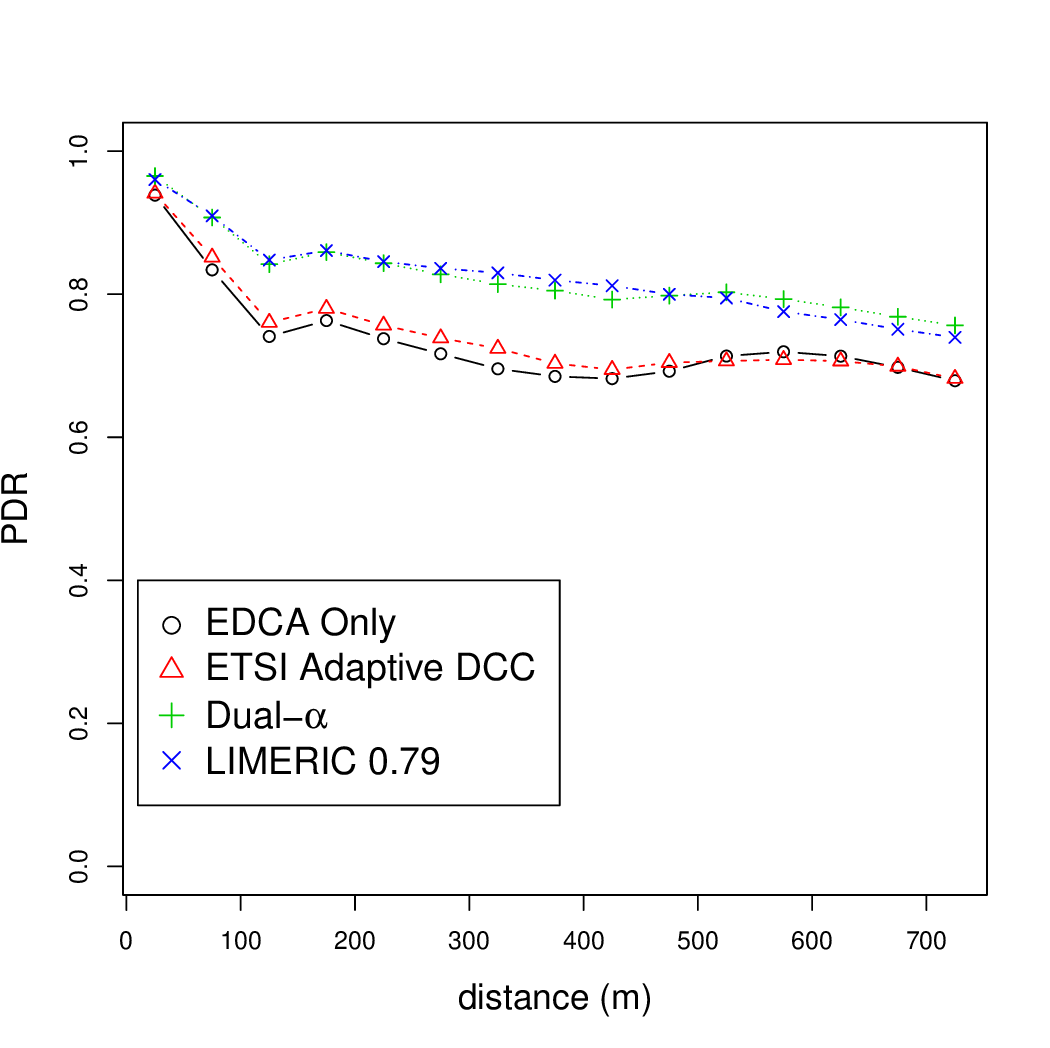}
	\caption{Packet Delivery Ratio in the saturated zone (25s<$t$<45s)}
	\label{fig:pdrdistancetl}
\end{figure}

\begin{figure}[h]
	\centering
	\includegraphics[width=205px,clip,keepaspectratio]{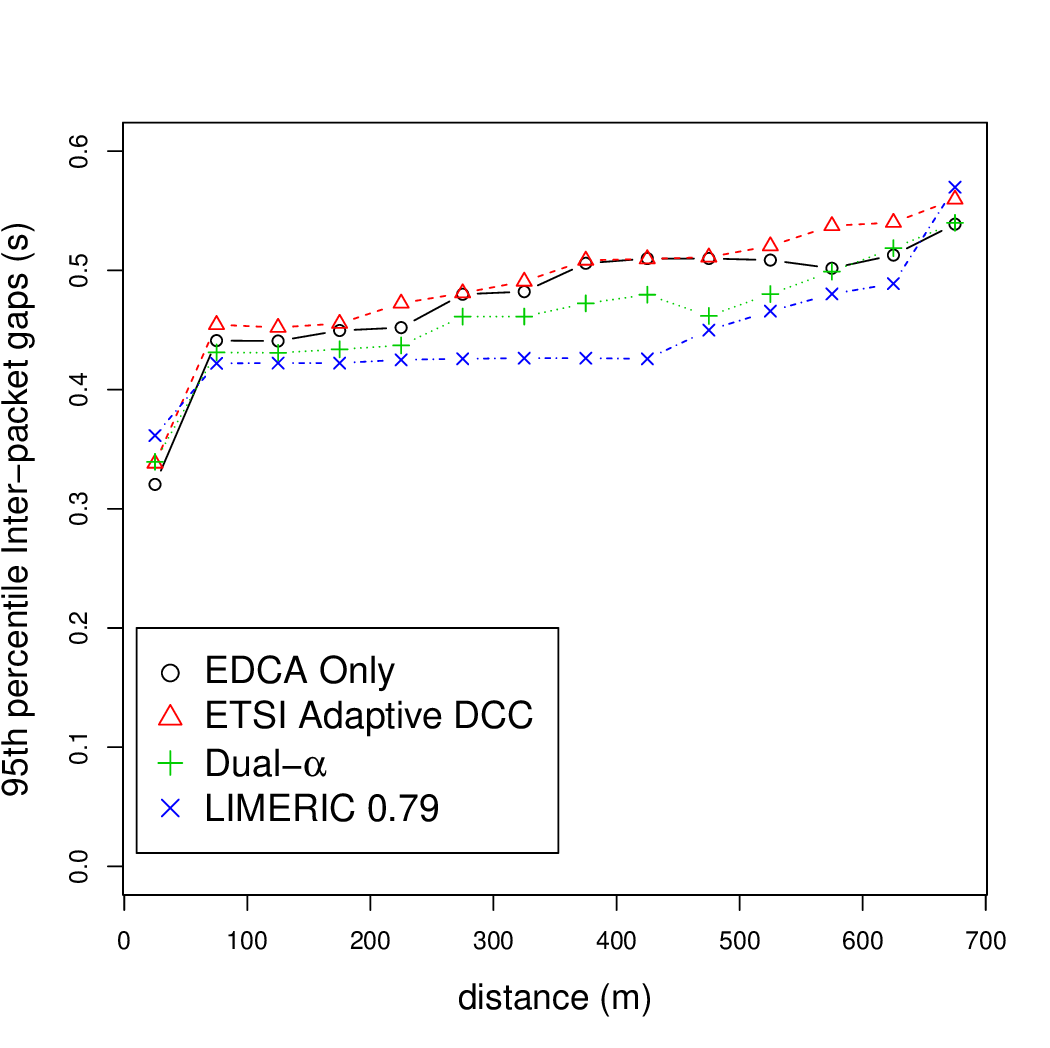}
	\caption{\textcolor{black}{95th percentile inter-packet gap in the saturated zone (25s<$t$<45s)}}
	\label{fig:gapsdistancetl}
\end{figure}

The effect of not being able to keep occupation under the 0.68 target is shown in Fig. \ref{fig:pdrdistancetl}, which shows PDR values for the time where the ETSI Adaptive Approach mechanism cannot keep occupation below its target. For the same extent of time, Dual-$\alpha$ and LIMERIC 0.79 manage to keep PDR over 80\%, while the other approaches are constantly below. There is also an effect on how often vehicles update their neighbors successfully, which is shown in Fig. \ref{fig:gapsdistancetl}, where 95th percentile inter-packet gaps for Dual-$\alpha$ and LIMERIC 0.79 stay below those for the ETSI Adaptive DCC even when $\delta$ in Dual-$\alpha$ and \textcolor{black}{LIMERIC 0.79 start restricting CAM generation earlier.}

There are several conclusions from these results: 1) maintaining channel conditions within a certain limit \mbox{($CBR<0.68$)} allows metrics such as PDR to stay in an acceptable region; 2) the parameter for $\alpha$ in the ETSI algorithm can cause the system to react slowly in transitory situations while the use of two values for $\alpha$ can help the system react faster and keep PDR levels and inter-packet gaps in a desirable level, as shown by the performance of Dual-$\alpha$, which does not allow the system to go over the target; \textcolor{black}{and 3) the combination of parameters for \mbox{LIMERIC 0.79} allows it to converge faster than the ETSI Adaptive DCC mechanism, but with a trade-off that comes in the form of a tighter limitation to the traffic a vehicle can send even when the medium is underutilized}.

\paragraph{Junction Scenario}
\label{fairness}
For this scenario, a small group of cars (25 \textcolor{black}{vehicle}s) traveling on a perpendicular two-lane road merge into a larger group of 300 \mbox{\textcolor{black}{vehicle}s} on a six-lane road. \textcolor{black}{We study the evolution of fairness in the sharing of the medium between cars in the large and small groups after merging}. All \textcolor{black}{vehicle}s generate CAM messages under the rules set by \cite{etsiCAM} with a Data Profile 2, and background traffic with a Data Profile 3 (less priority). \textcolor{black}{The background traffic, as above, is generated so that every \textcolor{black}{vehicle} has at least one message to transmit at any given moment, which results in achieving the maximum utilization of the medium allowed by the DCC mechanisms}. Measures are taken for 60 seconds, starting \textcolor{black}{just when both groups enter in range of each other and finishing after both groups have been merged and mixed}.  

\begin{figure}[h]
	\centering
	\includegraphics[width=205px,clip,keepaspectratio]{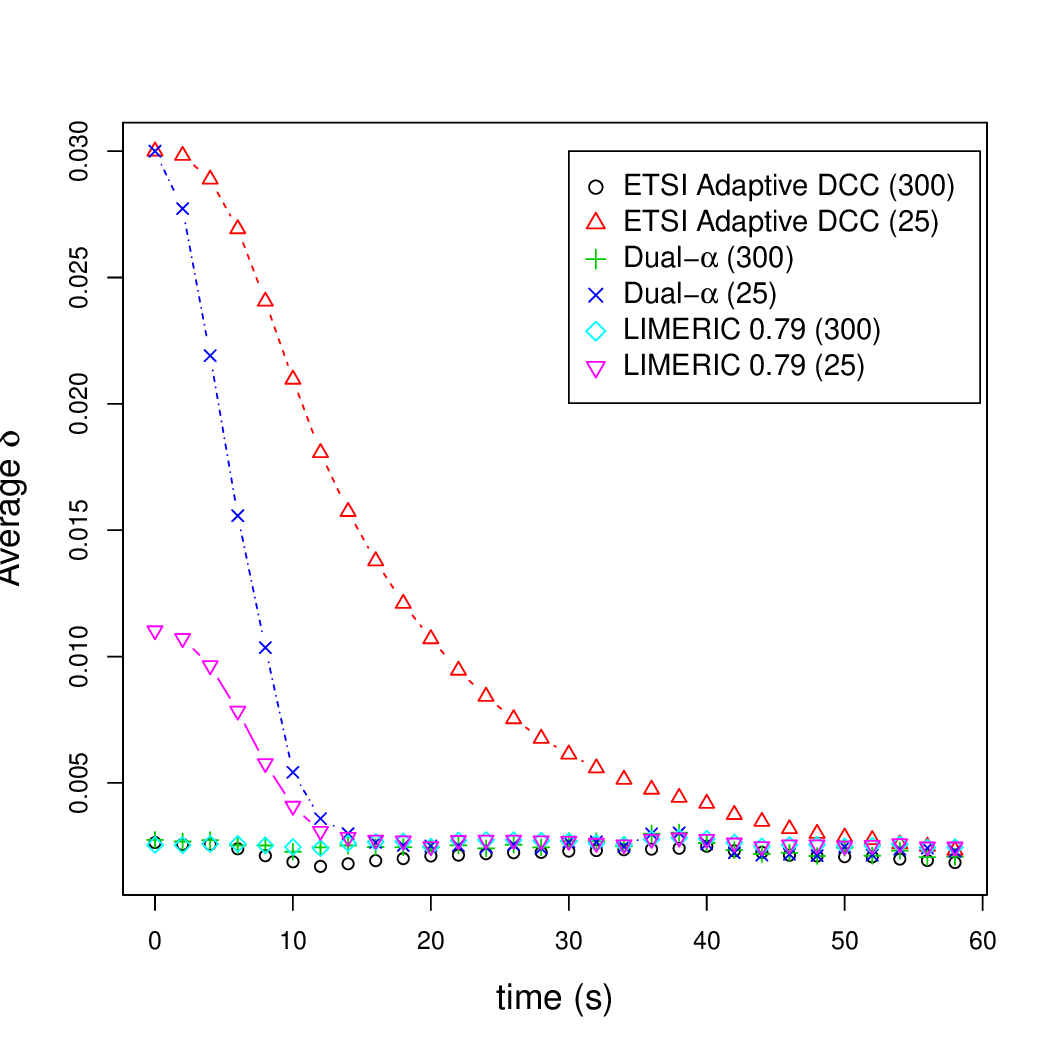}
	\caption{Evolution of $\delta$ between vehicles in one large and one small groups}
	\label{fig:deltas300x25}
\end{figure}
\begin{figure}[h]
	\centering
	\includegraphics[width=205px,clip,keepaspectratio]{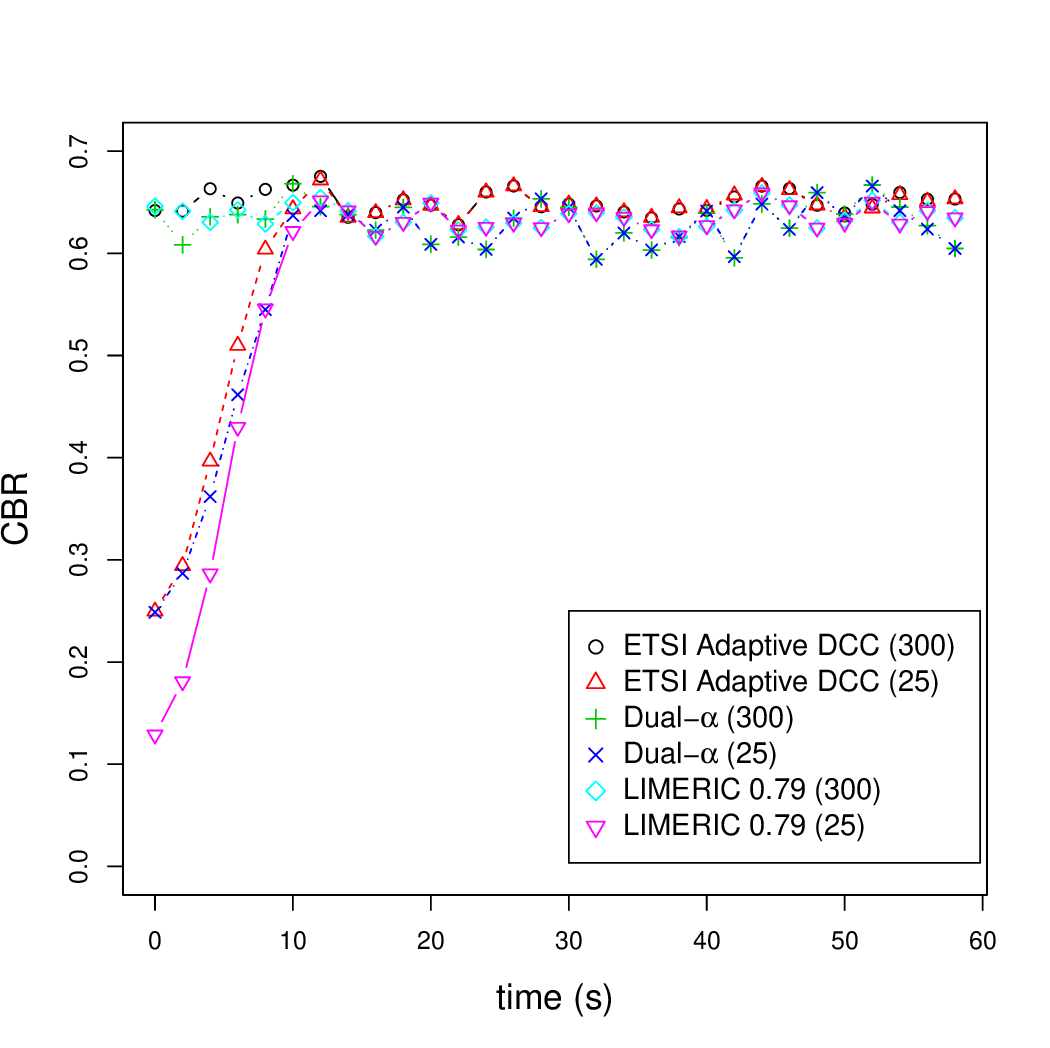}
	\caption{Evolution of channel occupation between vehicles in one large and one small groups}
	\label{fig:cbr300x25}
\end{figure}

\begin{figure}[h]
	\begin{subfigure}{.5\textwidth}
	\centering
	\includegraphics[width=205px,clip,keepaspectratio]{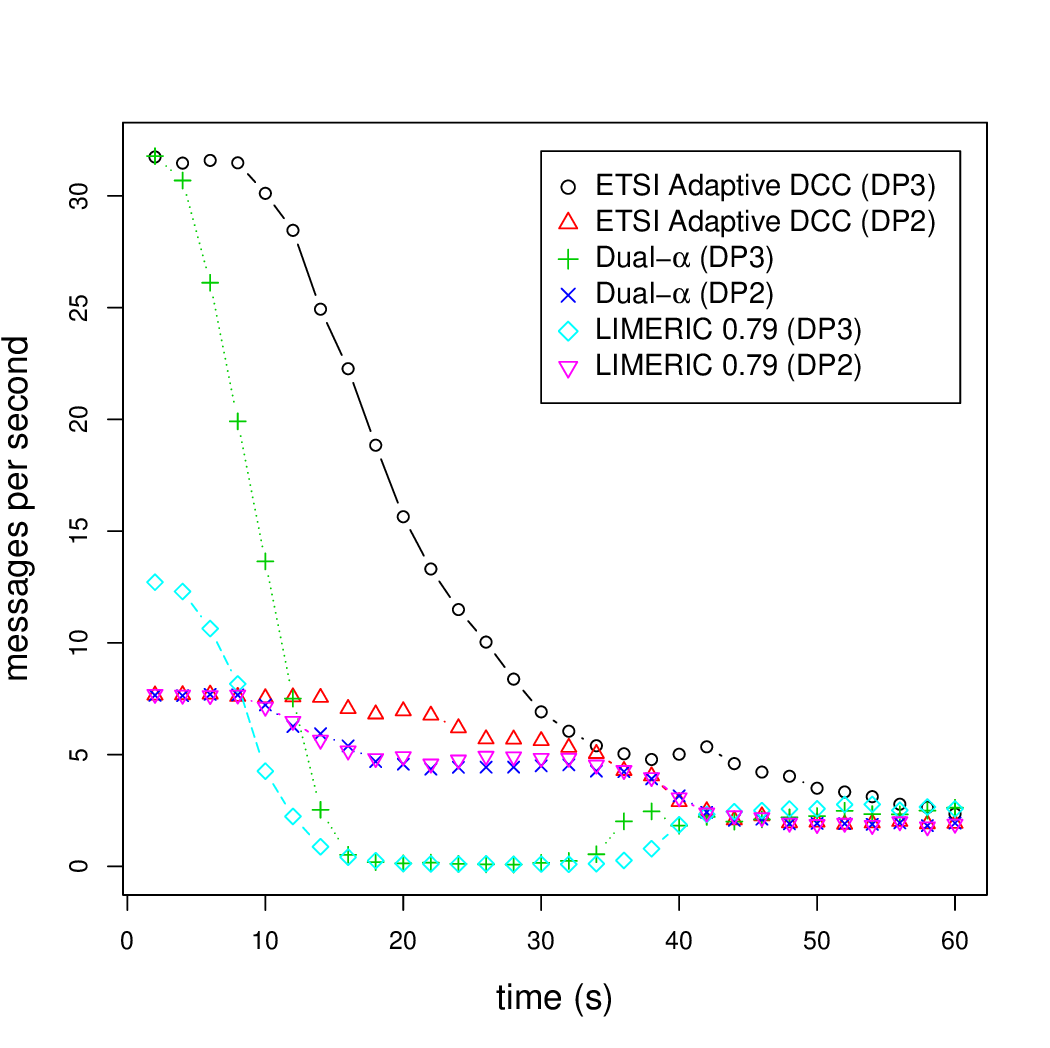}
	\caption{Small Group (25 vehicles)}
	\label{fig:smalltx}
	\end{subfigure}
	\begin{subfigure}{.5\textwidth}
	
	\centering
	\includegraphics[width=205px,clip,keepaspectratio]{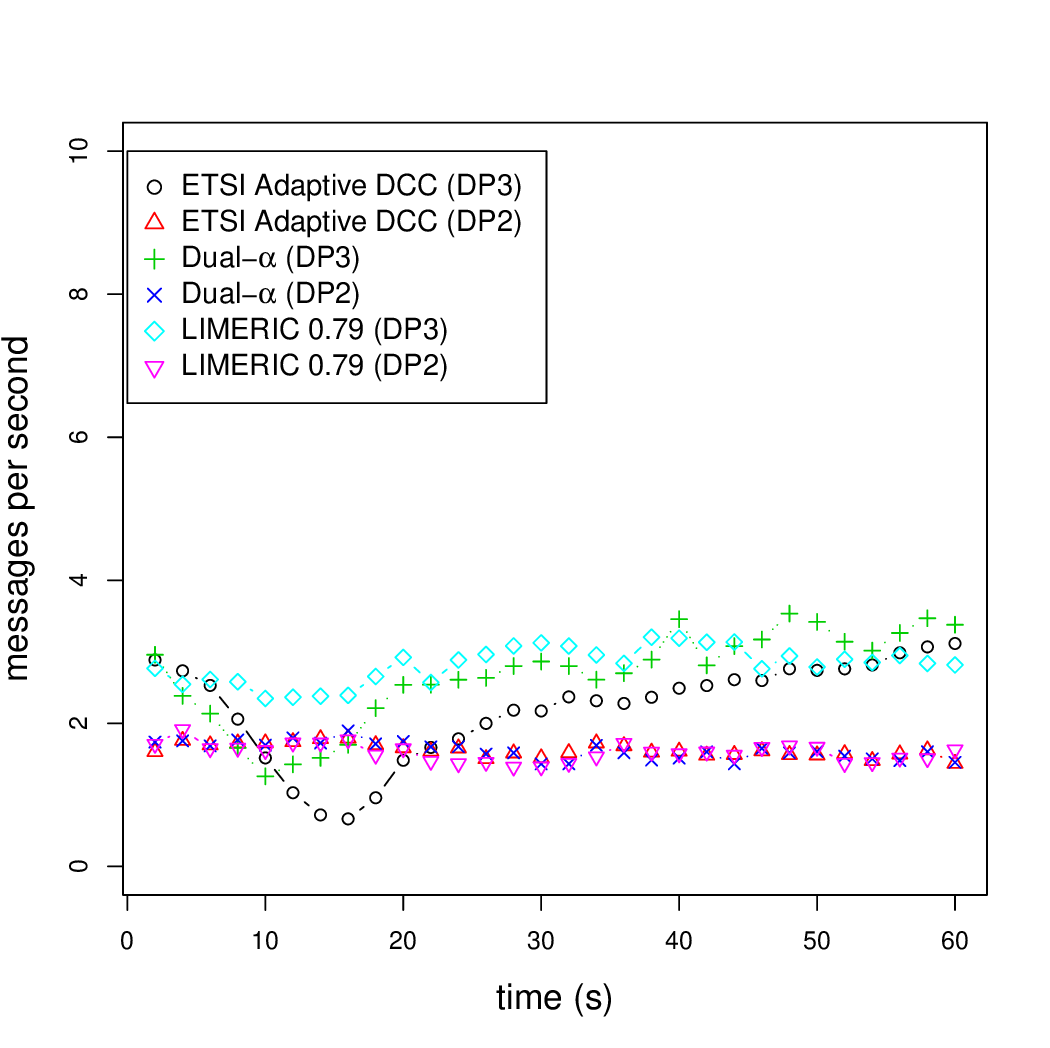}
	\caption{Large Group (300 vehicles)}
	\label{fig:largetx}
	\end{subfigure}
	\caption{Average DP2 and DP3 messages transmitted per second in the Junction Scenario}
	\label{fig:junctiontx}
\end{figure}

Before merging, with ETSI Adaptive DCC and \mbox{Dual-$\alpha$}, vehicles in the group of 25 \textcolor{black}{vehicle}s are converged to a $\delta$ that is close to $\delta_{max}$ in the parameters for the ETSI DCC Adaptive Approach mechanism, which allows them to transmit at 40 Hz, essentially transmitting every message they generate. With LIMERIC 0.79, the achieved $\delta$ is 37\% of $\delta_{max}$, due to the underutilization of the medium when using this mechanism at low densities. Vehicles in the large group have \textcolor{black}{values of $\delta$ which correspond to the convergence point for each mechanism}. All cars in the small group start hearing messages from the large group ($t=0$), which increases the channel occupation they measure and prompting them to lower their $\delta$ accordingly. Fig. \ref{fig:deltas300x25} shows that the ETSI Adaptive DCC algorithm takes approximately three times as much time to converge than DCC mechanisms with higher $\alpha$ values. \textcolor{black}{LIMERIC 0.79 converges as fast as Dual-$\alpha$, but it starts from a lower $\delta$ value that is reflected on the amount of traffic it allows to be transmitted by the small group}.  \textcolor{black}{Fig. \ref{fig:cbr300x25} shows that after approximately 10 seconds, both groups of vehicles are seeing the same medium utilization (i.e., they are in the same situation regarding medium sharing), while Fig. \ref{fig:deltas300x25} and Fig. \ref{fig:junctiontx} show how vehicles in the two groups are using the medium according to each DCC algorithm.} 

Differences in fairness can be seen in Fig. \ref{fig:junctiontx}, which represents \textcolor{black}{the average number of messages each \textcolor{black}{vehicle} in the small and large groups is allowed to transmit}. As seen in Fig. \ref{fig:deltas300x25}, $\delta$ for the small group descends slowly for the ETSI algorithm, and \textcolor{black}{vehicle}s are allowed to send not only their CAM traffic, but also a large quantity of background messages. In LIMERIC 0.79 and Dual-$\alpha$, after 13 seconds, $\delta$ is the same for vehicles in the small and large groups, but vehicles are using it differently. In the small group, vehicles send CAM traffic required by their dynamics (they are turning and incorporating to the main road), while background traffic cannot be sent because CAM traffic uses the complete share of the medium allocated to each vehicle. As time progresses, there is a decrease in vehicle dynamics, since vehicles in the small group are incorporated to the movement dynamics of the large group, so less CAM traffic is needed and background traffic starts to be sent. In the large group, at the beginning, the needed CAM traffic is lower (due to vehicle dynamics), and vehicles can also send background traffic. Note that, in LIMERIC 0.79 and Dual-$\alpha$, the total number of messages sent is the same in both groups (as expected because $\delta$ is the same) but it is used differently according to vehicle dynamics. On the other hand, in ETSI Adaptive DCC, vehicles in the small group are sending much more messages than vehicles in the large group. 

\begin{figure}[h]
	\centering
	\includegraphics[width=205px,clip,keepaspectratio]{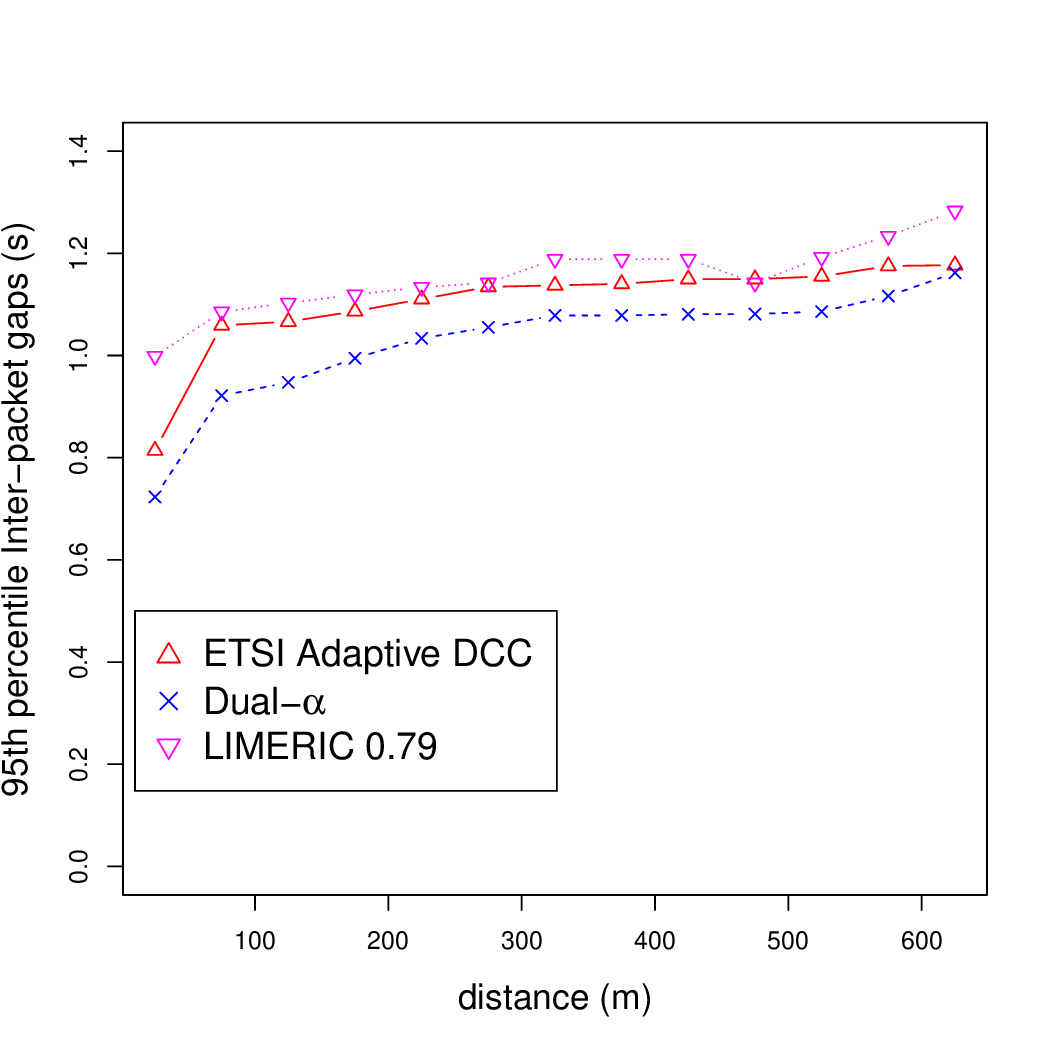}
	\caption{\textcolor{black}{95th percentile inter-packet gaps for the large group in the convergence zone (5s<$t$<25s)}}
	\label{fig:ipglargejunc}
\end{figure}

While the main purpose of having the small group converge to a $\delta$ that corresponds to the new conditions it faces (i.e., merging to a larger group) is to share the medium with fairness, it is also important to analyze the toll that convergence time takes on metrics such as inter-packet gap. Fig. \mbox{\ref{fig:ipglargejunc}} shows the 95th percentile inter-packet gaps for the large group of vehicles during twenty seconds starting five seconds after both groups sense each other. \textcolor{black}{For this transition zone, values for \mbox{Dual-$\alpha$} stay consistently lower from those of the ETSI Adaptive DCC and LIMERIC 0.79.}

Conclusions from this scenario can be summarized as follows: 1) the use of a high value for $\alpha$ improves fairness in transitory situations as can be seen in Fig. \ref{fig:deltas300x25}, where $\delta$ for the two groups converge three times as fast with LIMERIC 0.79 ($\alpha = 0.1$) and Dual-$\alpha$ ($\alpha_{high} = 0.1$) when compared to the ETSI Adaptive DCC ($\alpha = 0.016$); and 2) consequences of unfairness can be measured in metrics such as inter-packet gap, where the large group is affected by having to accommodate the excess of share of the medium allocated to the small group. 

\section{Conclusion}
\label{sec:conclusion}
We presented an analysis of the ETSI Adaptive Approach for decentralized congestion control (DCC) in vehicular communications and related algorithms. This work does not only consider transmission rate control at the access layer, but also at the facilities layer for awareness-related services. The ETSI Adaptive Approach works well over a wide range of vehicle densities, which validates the selection of parameters in the standard. Particularly, when compared with the absence of a DCC mechanism (i.e., using only EDCA), even in low network congestion scenarios, the use of the ETSI Adaptive Approach mechanism improves metrics such as Packet Delivery Ratio and how frequently remote \textcolor{black}{vehicle}s receive updates from other vehicles.

However, the parameters set by the ETSI Adaptive Approach \cite{etsiNewDcc} make the algorithm slow to converge in transitory scenarios. This results in situations in which channel occupation can be over the CBR target for significant amounts of time, and situations with unfair sharing of the medium among \textcolor{black}{vehicle}s. This is brought by the small value given to $\alpha$ in the ETSI standard. However, a greater value results in a convergence to a channel utilization that is further from the CBR target. A solution that has been tried in the literature is having a large value for $\alpha$ and increasing the CBR target (LIMERIC 0.79 \cite{LimericStability}), but we have shown that this combination of parameters \textcolor{black}{is very sensitive to the density of vehicles and the generated traffic, causing situations of underutilization and over-utilization of the medium.} 

Dual-$\alpha$ \cite{Letter}, using two values for $\alpha$, achieves better speed of convergence and fairness in transitory scenarios while still converging to a utilization close to the CBR target. We have shown that Dual-$\alpha$ performs similarly\textcolor{black}{, or even slightly better, than} the ETSI Adaptive Approach in scenarios with steady state conditions, while outperforming it in transitory situations. \textcolor{black}{This ability to deal faster with transitory scenarios can be important in many situations. For example, choosing the initial $\delta$ value in the ETSI DCC Adaptive Approach can be difficult. A high value means that, if many vehicles start at the same time (at the end of a sporting event, for example), there can be congestion and the DCC algorithm can be slow in reducing the load in the medium (similarly to the Junction scenario above). However, a small initial value of $\delta$ can mean that new vehicles joining the traffic are not announcing themselves as they should. This is not a problem in Dual-$\alpha$ because an initial high $\delta$ value will be quickly reduced, if needed, by the DCC algorithm.} 

\textcolor{black}{On the other hand, it must be considered that the ETSI DCC Adaptive approach is based on rigorous convergence properties studied theoretically in \cite{LimericOriginal}, while the use of two values of $\alpha$ in Dual-$\alpha$ potentially could create small oscillations in the allocated fraction of spectrum around the convergence point. So the advantages of Dual-$\alpha$ observed in practice should be carefully balanced against the robust theoretical properties of the ETSI DCC Adaptive approach.} 

Future work has to be performed in order to understand how DCC mechanisms can deal effectively with multiple types of traffic from a myriad of services that can coexist in the medium and serve their purpose in a fair manner.

\section*{Acknowledgment}
The authors thank reviewers for their valuable comments and suggestions. The quality of the reviews was very high and we believe that they have helped to improve the manuscript significantly.

\begin{IEEEbiography}[{\includegraphics[width=1in,height=1.25in,clip,keepaspectratio]{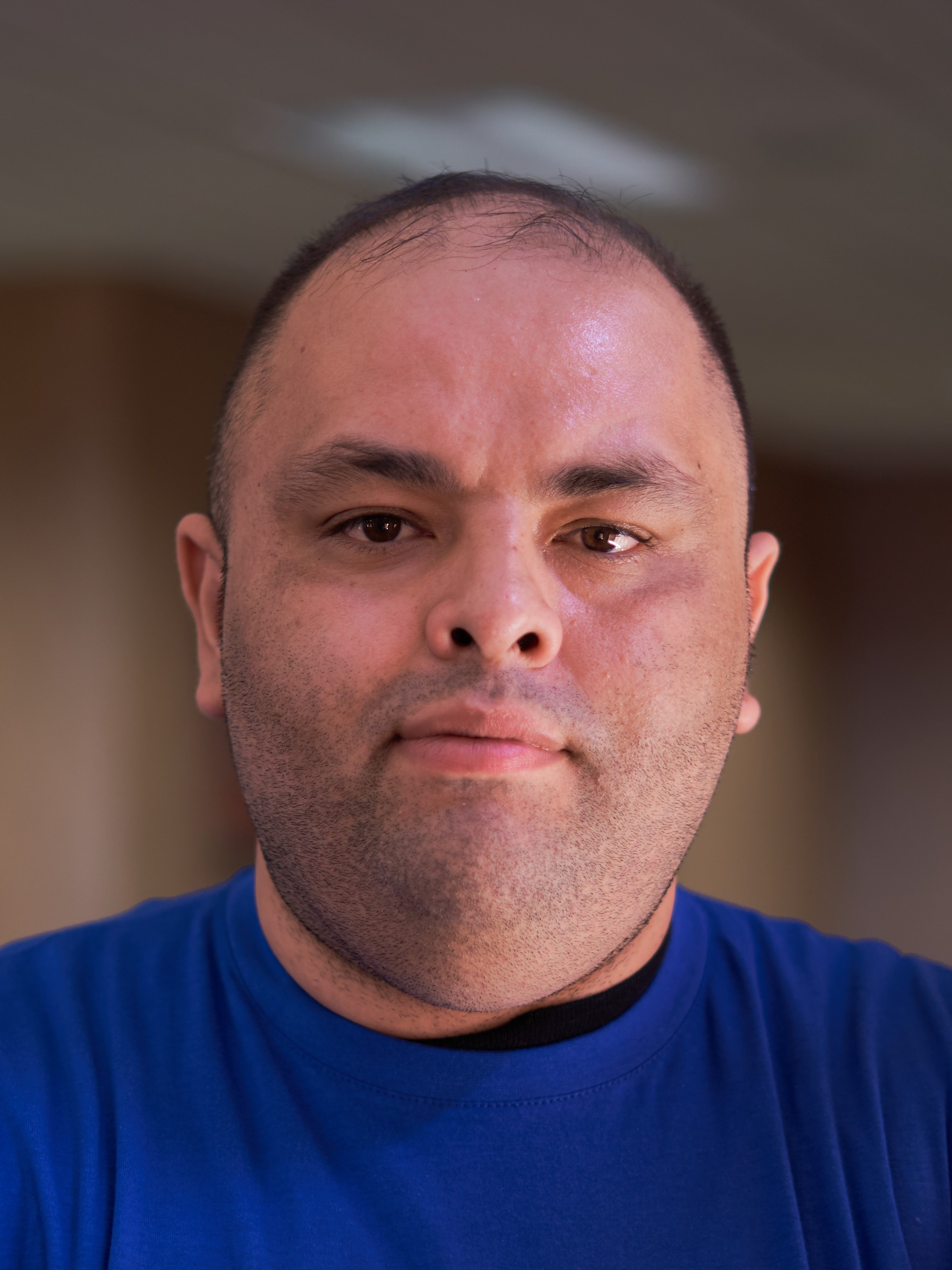}}]{Oscar Amador} is a Ph.D. student at the Telematics Engineering Department of University Carlos III of Madrid. He received a telematics engineering degree from Universidad Politecnica de Durango, in Mexico, and a M.S. degree in telematics engineering from University Carlos III of Madrid, in Spain, sponsored by a Fundacion Carolina scholarship. His research interests are focused on vehicular networking and intelligent transport systems. 
\end{IEEEbiography}

\begin{IEEEbiography}[{\includegraphics[width=1in,height=1.25in,clip,keepaspectratio]{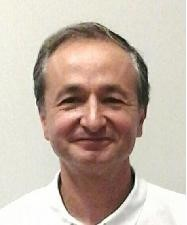}}]{Ignacio Soto} received a telecommunication engineering degree in 1993, and a Ph.D. in telecommunications in 2000, both from the University of Vigo, Spain. He was a research and teaching assistant in telematics engineering at the University of Valladolid from 1993 to 1999. He was associate professor at the Universidad Politécnica de Madrid from 2010 to 2012. In 1999 he joined University Carlos III of Madrid, where he currently works as an associate professor in telematics engineering. His research activities focus on security, mobility support in packet networks, multi-hop wireless networks, and vehicular networks.

\end{IEEEbiography}

\begin{IEEEbiography}[{\includegraphics[width=1in,height=1.25in,clip,keepaspectratio]{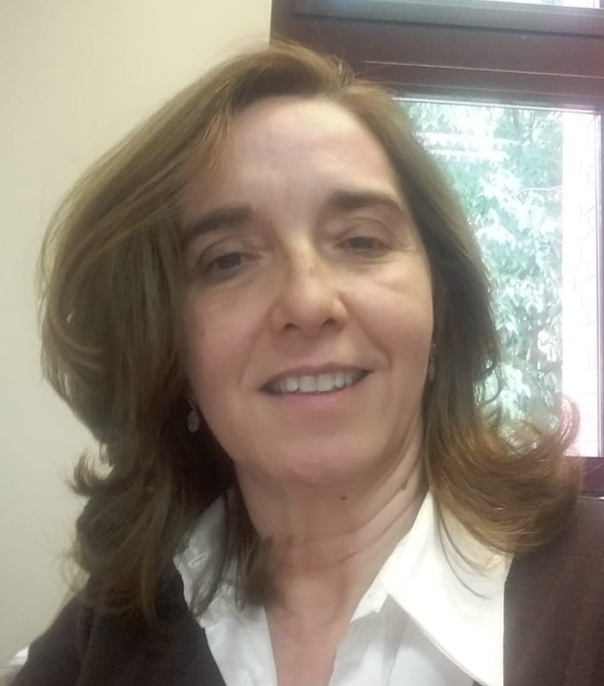}}]{Maria Calderon}  is an associate professor at the Telematics Engineering Department of University Carlos III of Madrid. She received a computer science engineering degree in 1991 and a Ph.D. degree in computer science in 1996, both from the Technical University of Madrid. She has published over 60 papers in the fields of advanced communications, reliable multi-cast protocols, programmable networks and IPv6 mobility. \textcolor{black}{She} has participated in research projects funded by the European programs FP7 and CELTIC, and different Spanish national programs. Her current work focuses on vehicular networks and mobile user services.
\end{IEEEbiography}

\begin{IEEEbiography}[{\includegraphics[width=1in,height=1.25in,clip,keepaspectratio]{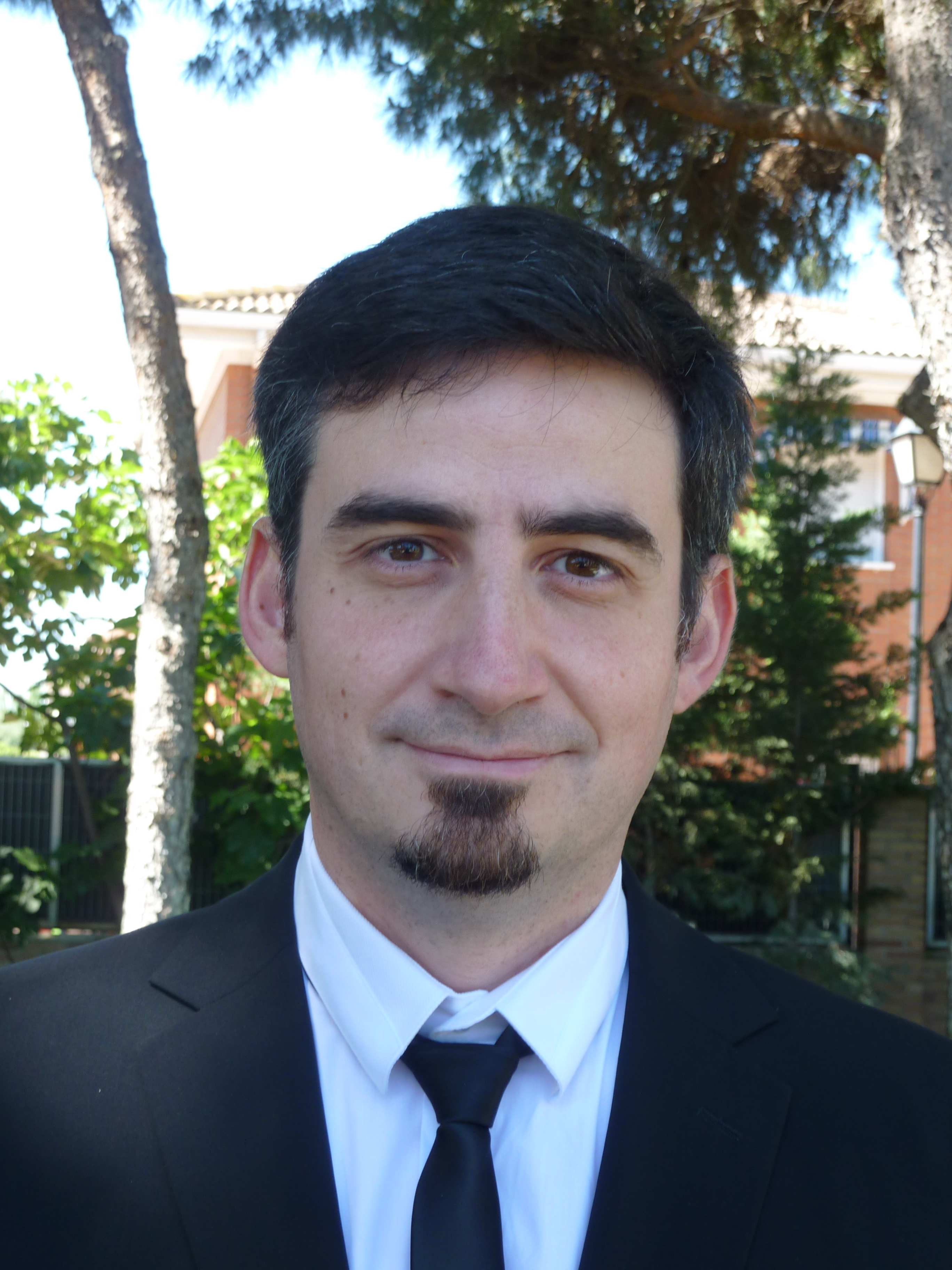}}]{Manuel Urueña}  received his M.Sc. degree in Computer Science from Universidad Politécnica de Madrid in 2001 and his Ph.D. degree in Telecommunications Technologies from Universidad Carlos III de Madrid in 2005. At present, he is an associate lecturer at Universidad Internacional de la Rioja. His research activities range from usable information security, through peer-to-peer and distributed systems, to vehicular ad hoc networks. He has been involved in several national and international research projects related with these topics, including the EU IST GCAP and the EU SEC INDECT projects.
\end{IEEEbiography}

\EOD

\end{document}